\documentclass[]{natureprintstyle}
\usepackage[normalem]{ulem}
\usepackage{graphicx}
\usepackage{longtable}
\usepackage{amssymb}
\usepackage{multirow}
\usepackage{savesym}
\usepackage{amsmath}
\savesymbol{iint}
\usepackage{txfonts}
\restoresymbol{TXF}{iint}
\usepackage{graphicx}
\usepackage{ulem} 
\usepackage{url}
\usepackage{threeparttable}
\usepackage{blindtext}
\usepackage{xcolor}
\usepackage{booktabs}
\usepackage[super,sort,comma]{natbib}
\usepackage{bibunits}
\usepackage[switch]{lineno}
\usepackage{subfigure}
\usepackage{etoolbox}
\usepackage[hidelinks, colorlinks=true, allcolors=black, pdfstartview=Fit, breaklinks=true]{hyperref}
\usepackage[hyphenbreaks]{breakurl}
\usepackage{overpic}

\makeatletter
\newcommand*{\newbibstartnumber}[1]{%
 \apptocmd{\thebibliography}{%
 \global\c@NAT@ctr #1\relax
 \addtocounter{NAT@ctr}{-1}%
 }{}{}%
}
\makeatother

\def\urlprefix {{\sc url: }}
\def\purple#1 {{\textcolor{purple}{#1}}\ }
\def\red#1 {{#1}}
\def\new#1 {{\bf #1 }}
\def\emph#1 {\textit{ #1 } }

\newcommand{\apj}{Astrophys. J.}

\newcommand{\apjs}{Astrophys. J. Supp.}
\newcommand{\araa}{Annu. Rev. Astron. Astrophys.}
\newcommand{\mnras}{Mon. Not. R. Astron. Soc.}
\newcommand{\apjl}{Astrophys. J. Let.}
\newcommand{\aap}{Astron. Astrophys.}
\newcommand{\aj}{Astron. J.}
\newcommand{\nat}{Nature}

\newcommand{\apss}{Ap\&SS}

\newcommand{\pasa}{Publ. Astron. Soc. Aust.}

\newcommand{\fermi}{{\em Fermi}\xspace}

\title{\bf A hyper flare of a weeks-old magnetar born from a binary-neutron-star merger}

\author
{B.-B. Zhang$^{1,2\ast}$,
Z. J. Zhang$^{1,2}$,
J.-H. Zou$^{1,2,3}$,
X. I. Wang$^{1,2}$,
Y.-H. Yang$^{1,2}$,
J.-S. Wang$^{4}$,
J. Yang$^{1,2}$, 
Z.-K. Liu$^{1,2}$,
Z.-K. Peng$^{1,2}$,
Y.-S. Yang$^{1,2}$,
Z.-H. Li$^{1,2}$,
Y.-C. Ma$^5$,
B. Zhang$^{6}$
\\
\\
\normalsize{$^{1}$School of Astronomy and Space Science, Nanjing University, Nanjing 210093, China }\\
\normalsize{$^{2}$Key Laboratory of Modern Astronomy and Astrophysics (Nanjing University), Ministry of Education, China}\\
\normalsize{$^{3}$College of Physics, Hebei Normal University, Shijiazhuang 050024, China} \\
\normalsize{$^{4}$Max-Planck-Institut f\"ur Kernphysik, Saupfercheckweg 1, D-69117 Heidelberg, Germany}\\
\normalsize{$^{5}$School of Artificial Intelligence, Nanjing University, Nanjing 210093, China }\\
\normalsize{$^{6}$Department of Physics and Astronomy, University of Nevada Las Vegas, NV 89154, USA}\\
\\
\normalsize{$^\ast$Corresponding author; E-mail: bbzhang@nju.edu.cn }
\\
}

\begin{document}

\begin{bibunit}[plainnat]

\maketitle

\begin{abstract}
Magnetars\cite{Duncan:1992ApJ.392L}, a population of isolated neutron stars with ultra-strong magnetic fields of $\sim 10^{14}-10^{15}$ G, have been increasingly accepted to explain a variety of astrophysical transients. A nascent millisecond-period magnetar can release its spin-down energy and power bright sources such as Gamma-ray Bursts\cite{Usov1992Natur} (GRBs) and their subsequent X-ray plateaus\cite{2006ApJ...642..354Z}, Super Luminous Supernovae (SLSNe) \cite{Kasen10,Woosley10}, 
and the fast X-ray transients such as CDF-S XT-2\cite{2019Natur.568..198X}. Magnetars with ages of $10^3-10^4$ years have been observed within the Milky Way Galaxy, which are found to power diverse transients with the expense of their magnetic energy, in the form of giant flares and repeated soft-$\gamma$-ray or hard X-ray bursts \citep{Thompson:1995MNRAS,Kaspi:2017ARA} and occasionally fast radio bursts (FRBs) \citep{CHIME/FRB:2020Natur,Bochenek2020}. Magnetar giant flares were also detected as disguised short GRBs from nearby galaxies \cite{Roberts2021Natur,2021Natur.589..211S,YangJun:2020ApJ}. Here we report the identification of a GRB as a hyper flare of magnetar in a nearby galaxy. The magnitude of the hyper flare is about one thousand times brighter than that of a typical magnetar giant flare. A significant $\sim 80$ millisecond period is detected in the decaying light curve. Interpreting this period as the rotation period and given a magnetic field strength typical for a young magnetar, the age of the magnetar is constrained to be only a few weeks. The non-detection of a (superluminous) supernova nor a GRB weeks before the event further constrains that the magnetar is likely born from an off-axis merger event of two neutron stars. Our finding bridges the gap between the hypothetical millisecond magnetars and the observed Galactic magnetars, and points toward a broader channel of magnetar-powered gamma-ray transients. 

\end{abstract}

\begin{figure*}
\begin{center}
\includegraphics[width = 0.4\textwidth]{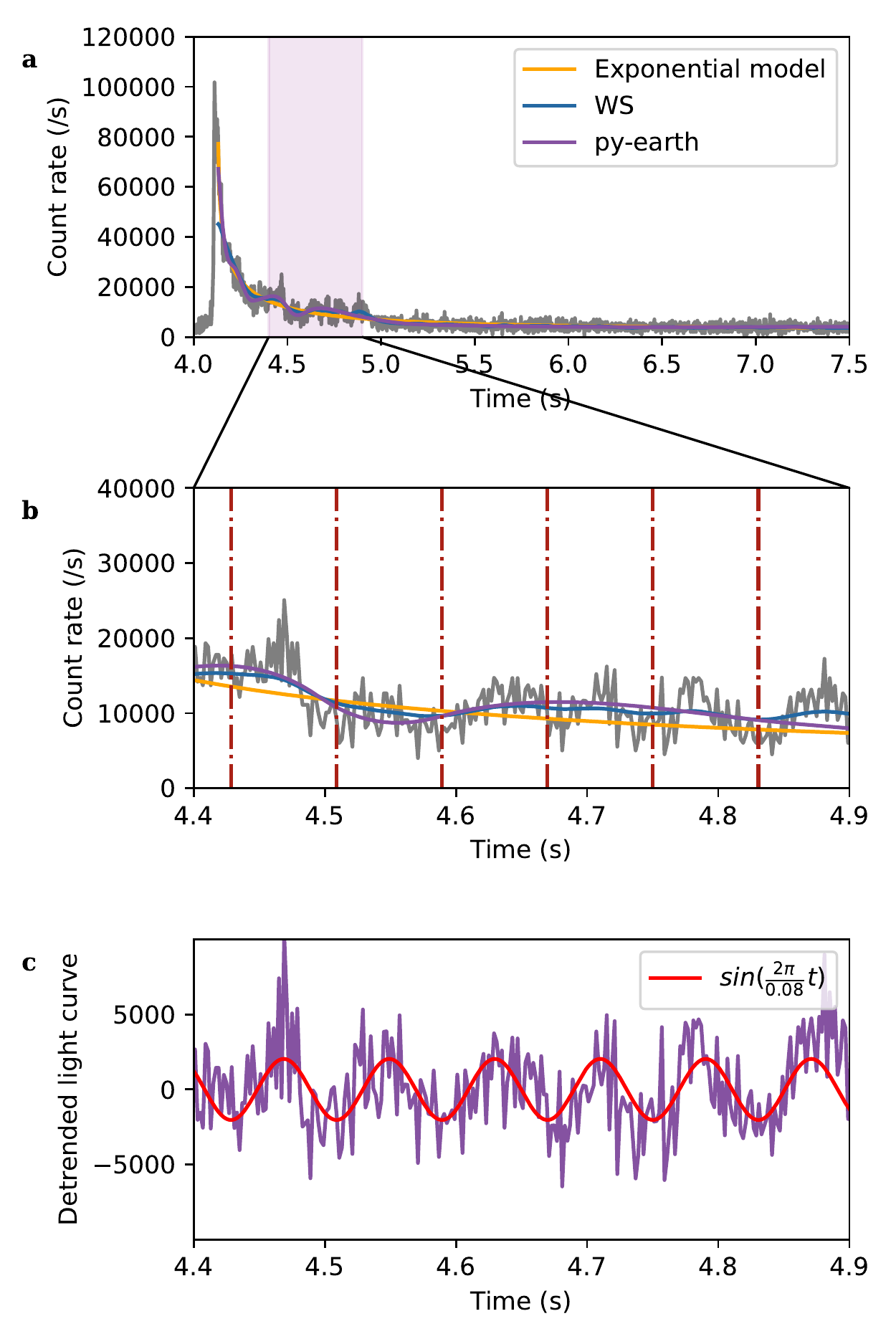}
\includegraphics[width = 0.3\textwidth]{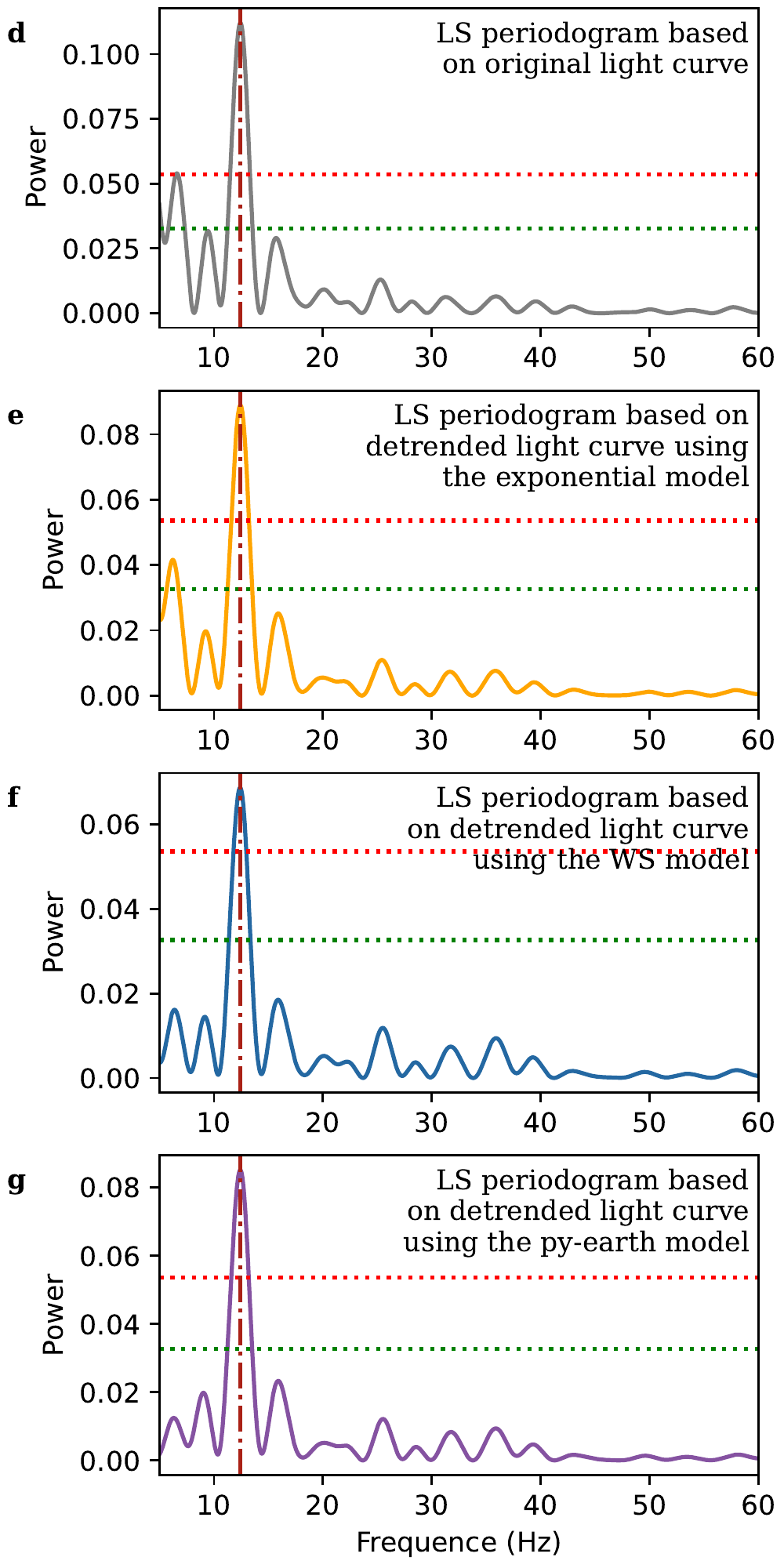}
\end{center}

\caption{
\noindent \textbf{Light curve and periodic signal detection of the hyper flare event, GRB 130310A.}
}
\label{fig1}
\end{figure*}
\begin{figure}
\begin{center}
\includegraphics[width = 0.5\textwidth]{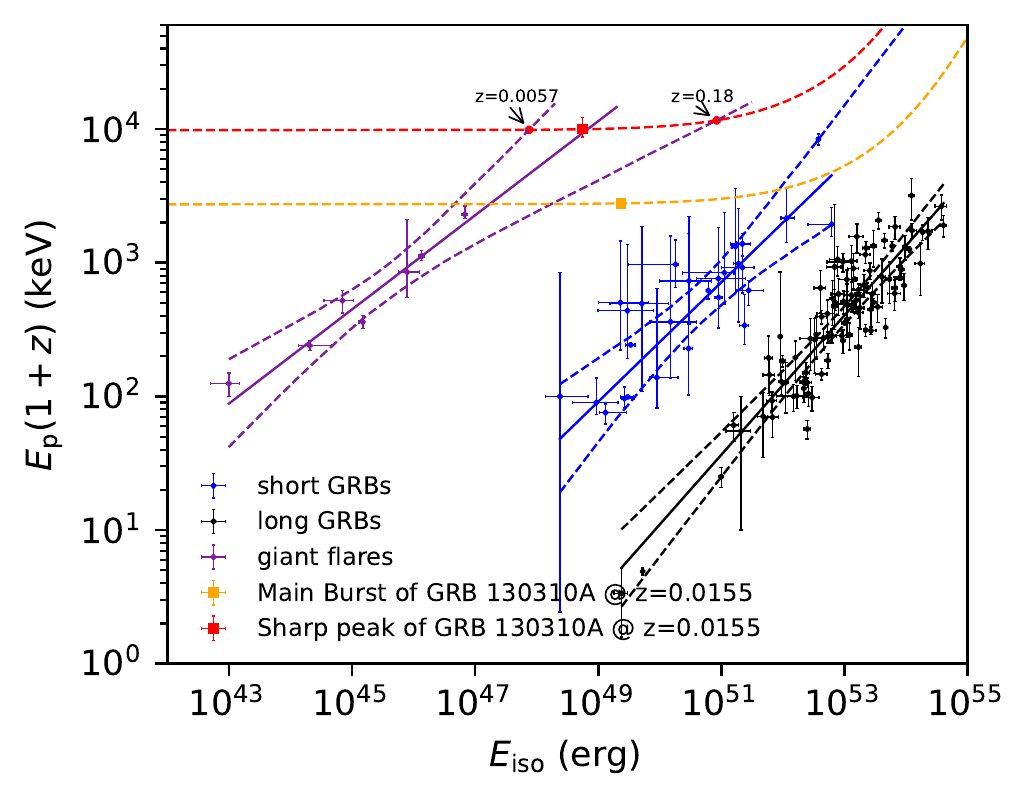}
\caption{
\noindent \textbf{ The $E_{\rm p}$ versus $E_{\rm iso}$ correlation diagram.} The upper left purple, middle blue and lower right black solid lines show the best-fit correlations for the MGF, short and long GRB populations, respectively. Dashed borderlines show the {3$\rm \sigma$} regions for each correlation. The red dashed line represents the locations of the sharp peak of GRB 130310A at different redshift values. The red square marks its location at a redshift of z = 0.0155, which is determined by its candidate host galaxy, g0927191-170053. The two red circles mark the allowed edge if the MGF folows the MGF-GRB track. All error bars represent 1$\sigma$ uncertainties.}
\label{fig2}
\end{center}
\end{figure}

Recent observations of the extragalactic magnetar giant flare (MGF)\cite{Roberts2021Natur,2021Natur.589..211S,YangJun:2020ApJ}, GRB 200415A, suggested that MGF GRBs follow 
the third track (other than long and short GRBs) in the rest-frame peak energy vs. isotropic energy ($E_{\rm p}-E_{\rm iso}$) plane\cite{YangJun:2020ApJ}. Motivated by this new track, we performed a systematic search in the Fermi/GBM GRB archives\cite{2020ApJ...893...46V}, aiming at finding additional MGF GRB cases. Our search started from looking for those GRBs with relatively high $E_{\rm p}$ and low $E_{\rm iso}$ (or low fluence if the redshift is not available), so they can be outliers of the short and long GRB populations in the $E_{\rm p}-E_{\rm iso}$ diagram. Our search quickly returned a strong candidate, GRB 130310A (Figure \ref{fig1}). {The burst consists of two main emission episodes: a precursor and a main burst, lasting a total of $\sim$ 4.2 seconds (see Methods and Extended Figure 1).} The precursor triggered Fermi/GBM\cite{2013GCN.14283....1X} at 20:09:41.503 on March $10^{th}$ 2013 UTC (hereafter $T_0$) and lasted for approximately 0.8 s. It is characterized by a thermal spectrum with $kT = 45.06^{+13.57}_{-5.69}$ keV (Methods). The main burst occurred at about $T_0+3.8$ s and presented a sharp peak followed by a series of erratic overlapping pulses and lasted for approximately 1.3 s {(Methods)}. Such a profile broadly resembles the previous observations of magnetar giant flares\cite{Mereghetti2005ApJ...628..938M}, which are characterized by a hard, sharp spike followed by a soft, long-lasting tail. The time-integrated spectrum of the main burst is characterized by a Cutoff Power Law (CPL) model with $E_{\rm p} = 2732.60_{-197.80}^{+229.75}$ keV and $\alpha = -1.10_{-0.01}^{+0.01}$. {Such an $E_p$ value is already significantly higher than most GRBs, making GRB 130310A, even with any assumed redshift between 0.0001 and 10, a distinct outlier from the long collapsar-type GRB track in the $E_{\rm p}-E_{\rm iso}$ diagram (Figure 2).} In consideration of its strong spectral evolution (Methods) as well as its spike+tail feature, we treat the 27-ms sharp peak (SP) (ranging from $4.108$ to $4.135$ s; as indicated in Extended Data Figure \ref{fig:t90lc}) as the characteristic emission of the event, and the following emission as its extended radiation. The time-integrated spectrum of the SP is best fitted by a CPL model parameterized by $\alpha = -1.19\pm 0.02$ and $E_{\rm p} = 9.8_{-1.2}^{+2.2}$ MeV, with a fluence of $9.88_{-0.82}^{+0.90}\times 10^{-6}$ erg cm$^{-2}$ (Methods). This suggests that the bulk Lorentz factor should be at least 
$\sim 430$ (Methods), being the highest among all observed MGFs observed so far \cite{Roberts2021Natur}.

Since there is no reportedly redshift measurement of GRB 130310A, we assigned $z$ as a free parameter ranging from $10^{-4}$ to $10$, and overplotted the corresponding values of the rest-frame $E_{\rm p}$ and $E_{\rm iso}$ of the SP onto the $E_{\rm p}$-$E_{\rm iso}$ diagram, as shown in Figure \ref{fig2}. Interestingly, due to its high $E_{\rm p}$, the event lies far away from typical long and short GRBs tracks. Instead, it is consistent with the third track for the MGF-GRB population. The redshift of the GRB is constrained in the range of [0.0079, 0.0595], assuming that the burst follows the MGF track. We further searched for the host galaxy (Methods) of the burst in the 6df Galaxy Survey (6dfgs) database\cite{2004MNRAS.355..747J,2009MNRAS.399..683J} within such a redshift range inside the overlapped region between the LAT error circle (Methods) and the IPN error box\cite{2013GCN.14284....1G}. Our search yields only one galaxy within that redshift range and error box, which is g0927191-170053 located at RA = $+09^h 27^m 19.70^s$ and DEC = $-17^{\circ}00^\prime53.0^{\prime \prime}$ with a redshift of 0.0155. At such a redshift, the isotropic energy (peak luminosity) of SP is $5.61_{-0.46}^{+0.51}\times 10^{48}$ erg ($2.30_{-0.25}^{+0.25}\times 10^{50}$ erg s$^{-1}$). With such a low energy and a high spectral peak, GRB 130310A consistently lies at the high-end of the MGF GRB population track (Figure 2). The total isotropic energy of the main burst is $2.37_{-0.18}^{+0.18}\times 10^{49}$ erg, which is around a thousand times higher than that of GRB 200415A. We hence consider it as a ``hyper" flare of a magnetar in this study.

The magnetar nature of GRB 130310A is further manifested by the period detection in the light curve. We employed the Lomb-Scargle (LS) method\cite{Lomb:1976,Scargle1982} to search for periodic signals in different time windows with different time scales in different energy ranges (Methods). As shown in Figure \ref{fig1}, we performed the LS calculation on the original light curves as well as the detrended light curves resulted from three different detrending methods. Our analysis yielded a significant periodic signal of 12.4 Hz with a confidence $>5 \sigma$ found in the 8-1000 keV light curve in the time region between 4.4 s and 4.9 s (Methods). Such a period is unprecedented in the magnetar observations and has never been observed in GRB events.

The period of $\sim$ 80 ms can be directly connected to the neutron star rotation.
Assuming magnetic dipole radiation, the spin-down rate, $\dot{P}$, can be estimated from the spin period, $P$, and its estimated surface magnetic field strength, $B_*$, as\cite{Shapiro1983}:

\begin{align}
\dot{P} &\simeq \frac{2\pi^2 R_{\rm s}^6 B_*^2}{3c^3IP} \nonumber\\ 
&= {1.94}\times10^{-8}\ \mathrm{s~s^{-1}} \bigg(\frac{P}{80\ \rm ms}\bigg)^{-1}\bigg(\frac{M}{2M_{\odot}}\bigg)^{-1}\bigg(\frac{B_{*}}{10^{15.5}~\rm G}\bigg)^{2}\bigg(\frac{R_{s}}{10^{6}~\rm cm}\bigg)^{6},
\end{align}
where $c$ is the speed of light, $M_{\odot}$ is the solar mass, $I \simeq 2MR_s^2/5$ is the moment of inertia, $M$ and $R_{\rm s}$ are the mass and surface radius of the magnetar, respectively. {We normalize $M$ to be $2M_\odot$ considering that the magnetar is likely born from a binary neutron star merger, as explained below.} Here we have adopted $Q = 10^n Q_n$ in cgs units. The magnetic field of a young megnetar can be as high as$\sim 10^{16}$ G\cite{Gourgou2016} and is normalized to $10^{15.5}$ (or 3.2$\times 10^{15}$) G in this study. The characteristic age of the magnetrar can be estimated as\cite{Shapiro1983}:

\begin{equation}
\tau_{c} \simeq \frac{P}{2\dot{P}} = {23.9}~\mathrm{\ days} \bigg(\frac{B_*}{10^{15.5}\rm\ G}\bigg)^{-2}\bigg(\frac{P}{80\rm\ ms}\bigg) ,
\end{equation}
which is about three weeks for nominal parameters. For simplicity, we further assume that there is about 50\% uncertainty in assigning the $B_*$ value, so the minimal and maximum age of the magnetar can be calculated as $\tau_{c,min}=11$ days and $\tau_{c,max}=56$ days respectively.

At age $\tau_c$, the spin-down luminosity is
\begin{align}
L_{\rm sd}=-\dot{E}_{\rm rot} &= -I\Omega\dot{\Omega} = 4\pi^2I\frac{\dot{P}}{P^3}= \frac{B_*^2R_{\rm s}^6\Omega^4}{6c^3} \nonumber\\
 &\simeq 2.35 \times 10^{42} \mathrm{erg\ s}^{-1} \bigg(\frac{P}{80\ \rm ms}\bigg)^{-4}\bigg(\frac{B_{*}}{10^{15.5}~\rm G}\bigg)^{2}\bigg(\frac{R_{s}}{10^{6}~\rm cm}\bigg)^{6}.
\end{align}
We note that such an $L_{\rm sd}$ is much smaller than the observed peak luminosity of GRB 130310A, suggesting that the hyper flare is not powered by the spindown of the magnetar, but is rather powered by significant magnetic energy dissipation in this early phase of a magnetar's life.

One may compare the GRB energy with the total energy available from the source. 
Considering a beaming factor of $f\sim 0.01$, the total energy of GRB 130310A may be estimated as $E_{\rm \gamma}=fE_{\rm iso}=2.37_{-0.18}^{+0.18}\times 10^{47} f_{-2}$ erg. The total spin energy of the magnetar at the epoch is
\begin{align}
E_{\rm rot} \simeq \frac{1}{2}I\Omega^2 \simeq {4.85}\times10^{48}~\mathrm{erg}\ \bigg(\frac{M}{2M_{\odot}}\bigg) 
\bigg(\frac{P}{80\rm\ ms}\bigg)^{-2}\bigg(\frac{R_{s}}{10^{6}~\rm cm}\bigg)^{6}, \label{eq:rotE}
\end{align}
and the total magnetic energy of the magnetar can be estimated as
\begin{align}
E_{B} \simeq \frac{1}{6}B_*^2R_s^3 \simeq 1.67 \times10^{48}~{\rm erg}\ \bigg(\frac{B_{*}}{10^{15.5}~\rm G}\bigg)^{2}\bigg(\frac{R_{s}}{10^{6}~\rm cm}\bigg)^{3}. \label{eq:magE} 
\end{align}
One can see that the GRB energy is within the energy budget of the magnetar if $f \ll 1$.
The hyper flare likely originates from the instantaneous release of the magnetic energy due to the dissipation of magnetic energy. The emission is likely highly beamed so that magnetar rotation can leave a periodic signal in the light curve. 
\begin{figure}
\begin{center}
\includegraphics[width = 0.52\textwidth]{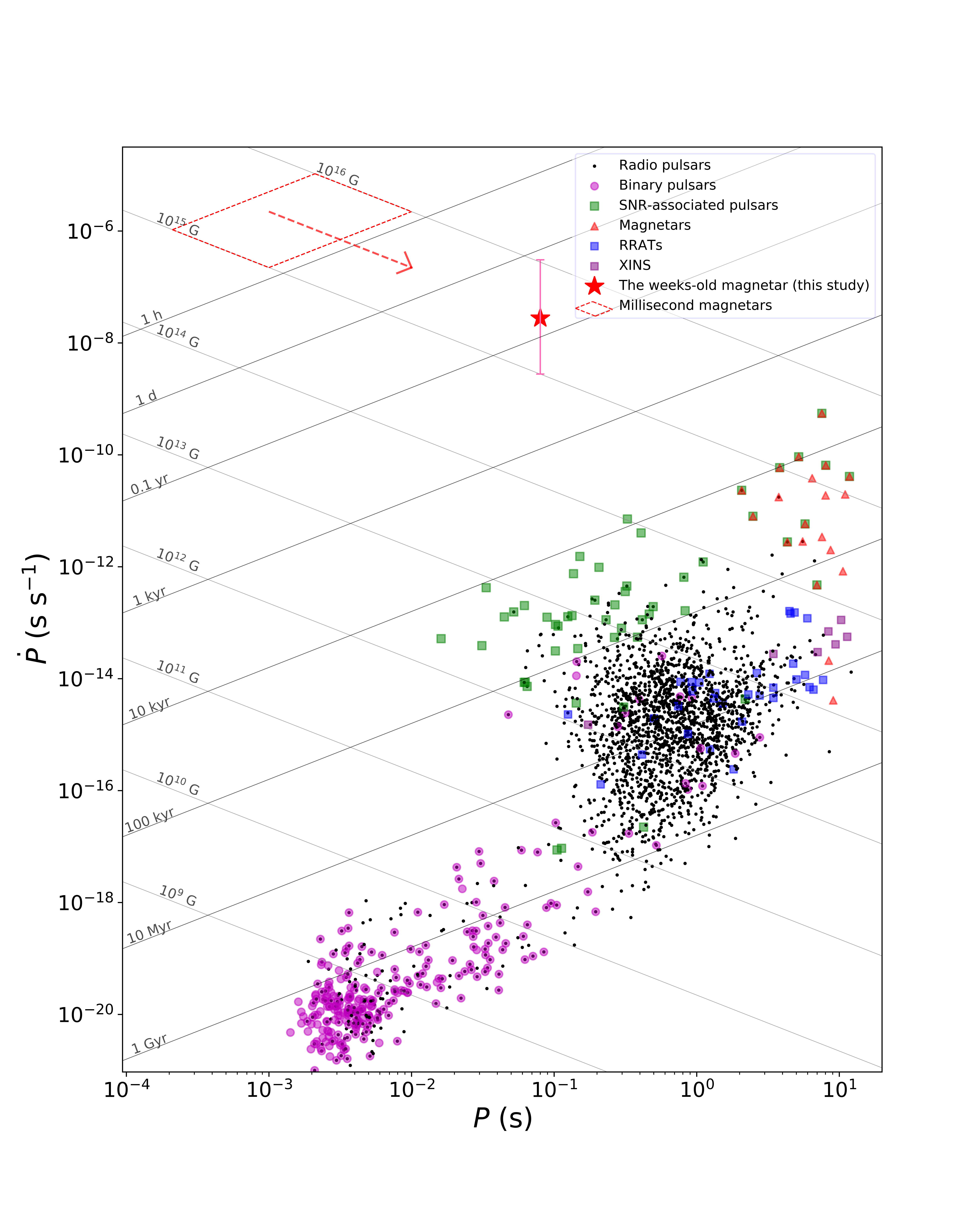}
\caption{
\noindent \textbf{${\rm P}-\dot{{\rm P}}$ diagram}. All known neutron stars collected in different categories in the ATNF Pulsar Catalogue\cite{ATNF} are plotted with different symbols as indicated in the top-right corner. The weeks-old magnetar in this study is overplotted with a red star. The hypothetical newborn millisecond magnetars with B$\sim 10^{15-16}$ G are located in the diamond region. The arrow indicates the evolution path from a newborn magentar to a weeks-old magnetar, assuming that $B$ does not decay significantly. }
\label{fig:ppdot}
\end{center}
\end{figure}

The estimated position of the magnetar powering GRB 130310A in the $P-\dot P$ diagram is presented in Figure \ref{fig:ppdot}. One can see that it bridges the hypothetical millisecond magnetars born in GRBs/SLSNe and Galactic magnetars. 

We searched for potential {GRBs} from the archival data as the possible progenitor of the magnetar. Our search window covers the time range between $T_{\rm 0}-\tau_{c,max}$ and $T_{\rm 0}-\tau_{c,min}$. We followed the burst search method described in Ref. \cite{Zou212021ApJ...923L..30Z}. The search yielded no significant signal in either triggered or untriggered GRBs samples around the location of GRB 130310A. The duty cycle, calculated by counting the time-span during which the GRB 130310A was in a good field of view (FOV) of Fermi/GBM, is {64.6}\%, suggesting that we had about {35.4}\% probability to miss the progenitor GRB event if there was any. Assuming a typical GRB spectrum parameterized by a CPL model with $\alpha=-1$ and $E_{\rm p}$ = 300 keV, the 10-s flux upper limit in 10-1000 keV at $T_{\rm 0}$-23.9 day is {$7.01\times10^{-8} \rm\,erg\,cm^{-2}\,s^{-1}$}, corresponding to a luminosity upper limit of $3.98\times10^{46}\rm\,erg\,s^{-1}$, which is $\sim 6000$ times fainter than the peak luminosity of GRB 130310A. {Our search suggests that a progenitor GRB must be significantly off-axis if it was not missed during the {35.4}\% off time. }

\begin{figure}
\begin{center}
\includegraphics[width = 0.52\textwidth]{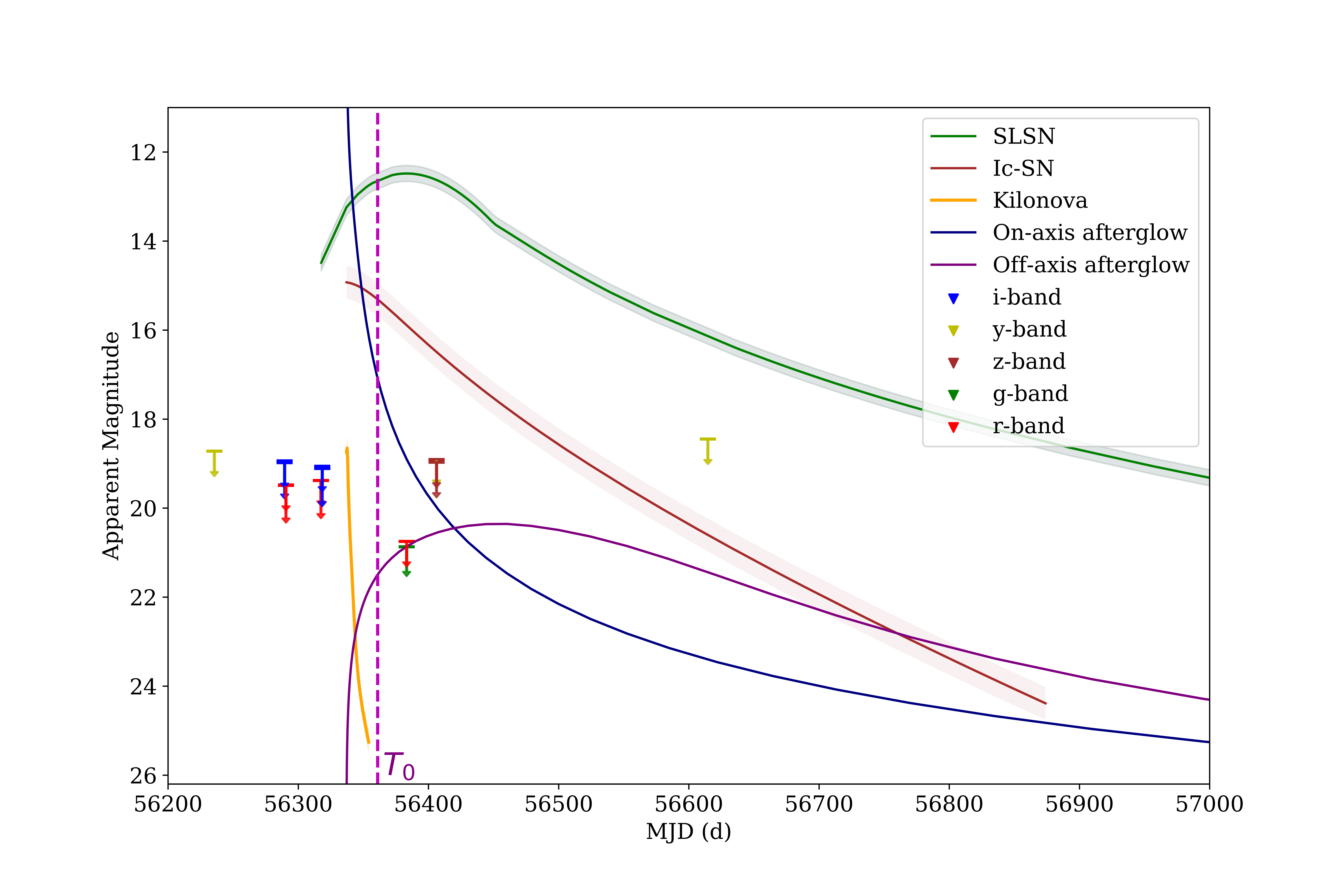}
\caption{
\noindent \textbf{Optical upper limits from the archival data and their constraints on optical emissions from the magnetar progenitor}. The arrows show the multi-band 3$\sigma$ upper limits for the non-detection of a point source transient in the skirt of the galaxy g0927191-170053 constrained from the PANN-STARR archival data (Methods). The solid green line shows a typical superluminous supernova light curve, assuming that it occurred at $T_0$ - 23.9 day at $\sim$ 70 Mpc. The brown solid line shows a typical type Ic supernova light curve assuming that it occurred at $T_0$ - 23.9 day at $\sim$ 70 Mpc. The orange, purple and navy curves stand for the kilonova, on-axis and off-axis optical afterglow light curves, respectively, for a GRB 170817A-like event located at $70$ Mpc.}
\label{fig:optical}
\end{center}
\end{figure}

We further utilize the archival data in the optical band to constrain the progenitor type of the magnetar. No supernova discovery was reported from the host galaxy within $T_0\pm$ 6 months. {The Panoramic Survey Telescope and Rapid Response System (Pan-STARRS)\cite{Panstarr_survey2016} largely covered the region of the magnetar from} {2012-06 to 2014-07}. We searched the PAN-STARRS data archive\cite{Flewelling2020_Panstarr} and found that there were {15 observations, from November 04, 2012 to November 18, 2013, with a total exposure time of $\sim$ 912 seconds}, covering the host galaxy region, which can place a series of $3\sigma$ non-detection upper limit for a point source around $T_0$ (Methods), as plotted in Figure \ref{fig:optical}. {Such upper limits can rule out the existence of a Type-Ic supernova associated with a long GRB or an SLSN at any time later than $T_0- \tau_{c,max} $ days (Methods). On the other hand, a kilonova or an off-axis afterglow of a GRB 170817A-like event at $\sim 70$ Mpc, can survive 
these constraints. Even for an on-beam short GRB, the afterglow emission may also avoid detection if it occurred in the gaps between PAN-STARRS observations. All these suggest that the progenitor of the magnetar is likely a neutron star merger event, with the off-axis viewing angle preferred because of its much higher probability. We note that the merger origin is also consistent with the fact that its host, g0927191-170053, is an old-type elliptical galaxy\cite{Campbell2014}.}

To date, the age of the magnetar is approximately $8.5$ years old. Using the total energy of $2.37\times10^{49}$ erg derived above and the standard afterglow parameters (Methods), we calculate the afterglow flux as $\sim$ $1.02\times10^{-12}$ / $3.04\times10^{-6}$ mJy in the X-ray/optical bands (Methods), which are way below the detection thresholds of the current missions/telescopes in operation. One possible way of detecting emission from the system may be fast radio bursts if the magnetar can survive until now. At $\sim$ 70 Mpc, 
an FRB with a typical luminosity of $10^{42}$ erg s$^{-1}$ should be readily detectable with most radio telescopes.

\setcounter{table}{0}

\setcounter{figure}{0}

\makeatletter
\renewcommand{\figurename}{Extended Data Fig.}
\renewcommand{\tablename}{Extended Data Table.}

\makeatother

\begin{methods}

\newbibstartnumber{30}

\vspace{1cm}

\section*{Fermi data analyses}

The hyper flare event, GRB 130310A, was a bright burst detected by several high-energy missions including \fermi/GBM \citep{2013GCN.14283....1X}, \fermi/LAT \citep{GCN14282}, Konus-Wind \citep{GCN14285}, and Suzaku WAM \citep{GCN14310}. We performed our analysis based on the time-tagged events (TTE) data collected by the Gamma-ray Monitor (GBM) onboard the Fermi satellite. The data were obtained from the Fermi Public Data Archive (\url{https://heasarc.gsfc.nasa.gov/FTP/fermi/data/}). We use the data of sodium iodide (NaI) detectors n9 \& na and germanium oxide (BGO) detector b1, which have the smallest angular separations with respect to the location of GRB 130310A and the highest brightness. We also utilized the Large Area Telescope (LAT) data covering 30 MeV to 100 MeV (LLE) and 100 MeV$-$300 GeV energy ranges. The standard data reduction procedure was described in \citep{Zhang:2011ApJ730,Zhang:2016ApJ816,Zhang:2018NatAs}. The key results of our analysis are outlined below.

\textit{Light curves and duration}: Light curves in different energy ranges are extracted from the TTE data following the procedures described in Ref\cite{Zhang:2018NatAs}. Those light curves are searched for possible periodic signals. The multi-wavelength light curves with typical energy ranges are presented in Extended Data Figure \ref{edf2}. The burst duration ($T_{\rm 90}$) is calculated in the standard range of 8 keV- 1 MeV. As shown in Extended Data Figure \ref{fig:t90lc} and listed in Extended Data Table \ref{tab:para}, the $T_{\rm 90}$, defined by the time interval during which 90\% of total counts are detected, is $\approx$ 1.3 s for the precursor and is $\approx$ $2.93~\rm s$ for the main burst.

\textit{Time lag}: 
Time lag measures the difference of photon arrival time in different energies\cite{2000ApJ...534..248N} and is often used as a probe to infer the emission size. We utilized the cross-correlation function (CCF)\cite{2000ApJ...534..248N,2010ApJ...711.1073U} to calculate the time lags of light curves among different energy bands between $T_{0}$ + $4.1 \rm $ s and $T_{0}$ + $ 4.6 \rm s$, following the method described in ref.\cite {2012ApJ...748..132Z}. As shown in Extended Data Figure \ref{edf2}, no significant lag is detected for the light curves in different energy bands. The zero-lag result suggests that the emission region of GRB 130310A is small, consistent with the speculation that it is an MGF GRB.

\textit{Amplitude parameter}: The amplitude parameter $f \equiv \frac{F_{\rm P}}{F_{\rm B}}$ is defined as the ratio between the peak flux $F_{\rm p}$ and background flux $F_{\rm B}$ (Ref. \cite{Lv14}). A small $f$ often suggests that the signal is more likely affected by the "tip-of-iceberg" effect. The $f$ parameter of the sharp peak of GRB 130310A is $11.22\pm 1.32$, which is the highest among all the GRBs (Extended Data Figure \ref{fig:f_para}).

\textit{Spectral analysis}: For the main burst ranging from $T_0 + 3.85 s$ to $T_0 + 5.25 s$, we performed both time-integrated and time-resolved spectral analyses using the data from the two NaI and one BGO detectors as mentioned ahead. The time slices for time-resolved analyses are obtained according to the light curve brightness profile in such a way that the photon count of each spectral channel is greater than twenty. All those intervals are listed in Extended Data Table \ref{exttab2}. The spectral files, including the total observed count spectra, the background spectra, and the detector response matrices (DRMs), are extracted from the event files using the same method as described in Ref. \cite{Yang:2020ApJ...899...60Y}. Four frequently used spectral models, namely Band function (Band), Black body (BB), simple power-law (PL), and cutoff power-law (CPL), are employed to fit the observed spectra. In particular, the CPL model is defined as \citep{Yu:2016A&A588}

\begin{equation}
	N(E) = AE^{\alpha} {\rm exp} [-(\alpha +2 )E/E_p ],
\end{equation} 
where $\alpha$ is the power-law photon index, $E_p$ is the peak energy of the fitted spectrum in units of keV, and A is the normalization factor. 

The time-integrated spectrum of the main burst is characterized by a CPL model with $E_p = 2454.34_{-189.19}^{+199.78}$ keV and $\alpha = -1.05_{-0.01}^{+0.01}$. We noticed that such an $E_p$ value is already significantly higher than that of most typical GRBs. 

The time-integrated spectrum of the shart peak alone is characterized by a CPL model with $E_p = 9802.23_{-1220.55}^{+2234.88}$ keV and $\alpha = -1.19\pm 0.02$. We noticed that such an $E_p$ value is already significantly higher than that of most typical GRBs. The total fluence of the first peak is $9.88_{-0.82}^{+0.90}\times 10^{-6}$ erg cm$^{-2}$.

By comparing the goodness of the fits in each time slice of the main burst, we found that the CPL model is the best model that adequately describes the observed data with the lowest Bayesian Information Criteria (BIC)\footnote{BIC\citep{Schwarz:1978} is defined as 
$ {\rm BIC} = -2{\rm ln} \mathcal{L}+k\ln n$, where $\mathcal{L}$ is the maximum likelihood, k is the number of parameters of the model, and N is the number of data points used in the fit.}, we thus only employ the CPL model to perform the time-resolved spectral fitting. 

Interestingly, the time-resolved spectral analysis suggests even more extreme values of $E_{\rm p}$. With significant spectral evolution within the $T_{\rm 90}$ interval, Ep reaches at $\sim 10^{+1.5}_{-3.2}$ MeV around the peak region of the light curve. As shown in Extended Data Figure \ref{edf3}, we note that $\alpha$ and $E_{\rm p}$ evolution shows an intensity tracking behavior as suggested by \cite{Wheaton1973,Golenetskii1983,Lu2012}.

\textit{High-energy photons}: We extracted the LAT data in the energy range of 100 MeV$-$300 GeV from $T_0$ to $T_0 +800$ s. The region of interest (ROI) is a square with the width of $10^{\circ}$ and the center at the position RA, Dec = 142.34, -17.23 (J2000). Using the standard Fermitools \citep{2020ApJS..247...33A}, and with a cut on zenith angle at $100^{\circ}$ and the instrument response function set P8R3\_SOURCE\_V3, an unbinned likelihood point source analysis is performed. The GRB spectral model is assumed to be \textit{Powerlaw2} \footnote{\url{https://fermi.gsfc.nasa.gov/ssc/data/analysis/scitools/source_models.html}} with $E_{min} = 20$ and $E_{max} = 20000$. Consequently, the spectral index is fitted to be $-1.9\pm 0.1$. Finally, we obtained 20 photons with different probabilities originating from GRB 130310A. Those high-energy photon events are plotted in Extended Data Figure \ref{edf2}a \& \ref{edf2}b.

\section*{Amati relation}

The correlation between the GRB isotropic energy $E_{\rm\gamma,iso}$ and the rest-frame peak energy $E_{\rm p,z}=(1+z)E_{\rm p}$ is commonly refereed to as Amati relation\cite{Amati:2002}, and can be written as ${\rm log~}E_{\rm p,z}=a+b{\rm log~}E_{\rm \gamma,iso}$. Typically, long and short GRBs, as well as MGF GRBs follow different tracks in the diagram\cite{Yang:2020ApJ...899...60Y}. We plotted the $E_{\rm p,z}-E_{\rm\gamma,iso}$ diagram (Figure \ref{fig2}a) using GRB samples with known redshift from Ref.\cite{Amati:2002} and Ref.\cite{Zhang:2009ApJ}. Utilizing the MCMC method, the optimal fitting parameters and 3$\sigma$ uncertainties are constrained. We obtain $a = -25.01^{+3.91}_{-4.25}$ and $b = 0.52^{+0.08}_{-0.07}$ for long GRBs, $a = -19.92^{+9.18}_{-7.60}$ and $b = 0.45^{+0.15}_{-0.18}$ for short GRBs, and $a = -13.20^{+7.82}_{-5.62}$ and $b = 0.35^{+0.12}_{-0.17}$ for MGF GRBs. The hyper flare event, GRB 130310A, is a significant outlier from both the short and long GRB tracks, but is consistent with being an MGF GRB. Assuming that it follows the MGF track, the redshift of GRB 130310A can be constrained within the range from 0.0079 to 0.0595.

\section*{Periodicity Measurement }
To maximize the signal detection, we combined all the photon events from the three detectors n9, na, and nb to search for periodic signals. The light curves in different energy bands are all binned with 0.0005s, which ensure sufficient resolution for the periodicity search. We searched possible periodic signals in the light curve of the main burst by using the Lomb-Scargle (LS) periodograms method. In addition to the original light curves, our search also considers the detrended light curves, with the latter aiming to remove the effect of global temporal evolution of the event. The following three detrending models are employed and fitted to the search phase of the light curve:

\begin{enumerate}
 \item Exponential model $f(t)= ae^{-b(t)^{d}}+c$, indicated as yellow solid line in Figure \ref{fig1}. 
 \item Whittaker Smooth\cite{Whittaker1922,Eilers2003} (WS) model, indicated as blue solid line in Figure \ref{fig1}. WS is a fast and wildly used smooth approach, which can well fit the trend of the light curve and retain the periodic signal by giving an appropriate smoothing parameter.
 \item A model using the multivariate adaptive regression splines algorithm\cite{friedman1991}, coded as py-earth model and indicated as the purple solid line in Figure \ref{fig1}. The multivariate adaptive regression splines algorithm is a non-parametric regression method that builds multiple linear regression models on each different partition, which can efficiently catch the global temporal features of the light curve. 
\end{enumerate}

The LS periodogram calculated based on the original light curve as well as the de-trended light curves by the above three methods are shown in the right column of Figure 1. A significant periodic signal of 12.4 Hz, corresponding to a period of 0.08s, is detected in all four situations. The strongest detection of the signal lies in a energy range of 50-300 KeV, and time range of 4.4-4.9 s. The confidence level of the detection in all cases exceeds 5$\sigma$, with a false alarm probability (FAP)\cite{2008MNRAS.385.1279B,VanderPlas2018} of {$\leq 5.7\times 10^{-7}$}. 

The periodic signal as well as its detectable range are also verified by applying the weighted wavelet
Z-transform) to the detrended data, as shown and Extended Data Figure 7.

The lack of detection of the periodic signal after 4.9 s in the extended tail is not unexpected as the observed flux can become less modulated by the underlying periodic signal when it decreases exponentially and emerges to the background level. This can be illustrated in Extended Data Figure 8, where we generated a series of Fast-Rising-Exponential-Decaying (FRED) shape light curves with injected noise similar to that of GRB 130310A, as well as a periodic $f=12$ Hz signal, $S_{\rm p}$, characterized by different amplitude $A$, time scale $\tau$, which is formulated by
\begin{equation}
 S_{\rm p}(t) = \frac{A}{{\rm exp}\big[(t-4.4)/\tau\big]} {\rm sin}(2\pi ft)
\end{equation}.

As shown in Extended Data Figure 8, our simulations indicate that the periodic signal can become undetectable after 4.9 s in the extended tail under certain configurations of the $S_{\rm p}(t) $.

\section*{Implication on the bulk Lorentz factor}

Following the derivation from Ref.\cite{2001ApJ...555..540L}, for a photon spectrum modelled as $fe^{-\alpha_{+}}$, in which $f$ is the normalization factor in unit of ph cm$^{-2}$ s$^{-1}$ keV$^{-1}$ and $\alpha_+$ is the inverse number of the photon index, $\alpha$, in correspondence with the definition in Eq. 6, the lower limit of the bulk Lorentz factor can be constrained by:

\begin{equation}
\gamma >\gamma_{\rm min}=\hat{\tau}^{1\over 2\alpha_++2}({E_{\rm max}\over m_ec^2}) ^{\alpha_+-1\over 2\alpha_++2}
\label{eq:gamma_lim}
\end{equation}

\begin{equation}
\hat{\tau}\equiv
{\eta (\alpha_+) \sigma_T d^2 (m_e c^2)^{-\alpha_++1}f\over c^2 \delta T (\alpha_+-1)},
\label{eq:tauhat}
\end{equation} 
where $E_{\rm max}$ is the photon with the highest energy, $\delta T$ is the minimal variability time scale constrained through the Bayesian block method, d is the distance of the burst, $\sigma_T$ is the Thomson cross-section and $\eta(\alpha_+)$ is the correction factor which reflects the averaging effect of the pair production cross section. The form of $\eta(\alpha_+)$ varies in the literature\cite{Svensson1987MNRAS.227..403S,Gupta2008MNRAS.384L..11G} and we adopt $\eta(\alpha_+)=(3/8)(1+\alpha_+)^{-1}$ as discussed in Ref.\cite{Gupta2008MNRAS.384L..11G}.

Observationally, the minimal variability time scale of GRB 130310A is 7.5 ms. The sharp peak whose spectrum is fitted as a CPL model parameterized by f=98.57, $\alpha_+$ = 1.19, and $E_{\rm max}\simeq E_{\rm p}=9.8$ MeV. By putting those numbers in Eq. 7, we can calculate the lower limit of the bulk Lorentz factor as $\gamma_{\rm min}\simeq$ 430.

\section*{Localization with Fermi/LAT and IPN}

By removing the source model of GRB 130310A from the fitted model, we construct the Test Statistics (TS) map by \textit{gttsmap}, and we localize GRB 130310A at RA = 142.52 and DEC = -17.16 with the maximum TS value. To obtain the confidence interval, we subtract the ${\rm TS} _{\rm max}$ from the TS map to build a Localization Test Statistics (LTS) map in Extended Data Figure \ref{fig:latloc}. The white, orange, green, and black ellipses denote the 68.3\%, 95.5\%, 99.7\%, and 99.9\% confidence areas, respectively. The LAT counterpart of the peak of the GBM light curve is located with RA = 144.50 and DEC = -19.02.

The InterPlanetary Network (IPN) reports a constraint of the burst location from a group of spacecraft equipped with gamma-ray burst (GRB) detectors\cite{2013GCN.14284....1G}. The overlapping region between the IPN error box and the $99$\% confidence of the LAT box is regarded as the most probable location region of the hyper flare event and is used for the host galaxy search. 

\section*{Host galaxy search}

Within the redshift range constrained by the Amati relation and spatial range of the overlapping IPN \& LAT error boxes, we performed a host galaxy search in the 6df Galaxy Survey (6dfgs) database\cite{2004MNRAS.355..747J,2009MNRAS.399..683J}, which records more than 100,000 galaxies in the nearby universe. Only one host galaxy, g0927191-170053 with RA = $+09^h 27^m 19.07^s$ and DEC = $-17^{\circ}00^\prime53.0^{\prime \prime}$ and redshift = 0.015492, was found. The galaxy is also cross-matched in the Two Micron All Sky Survey Extended Source (2MASS), as {09271905-1700528\cite{2011ApJ...737..103S}}. The location of the galaxy is marked in Extended Data Figure \ref{fig:latloc}.

\section*{Optical upper limits of the magnetar emission around T$_0$}

We search for all available archival optical sky-survey data in the hope of finding observations that can cover several months around T$_0$. Those observations can provide upper limits, if not detection, of a point source associated with the magnetar's optical counterpart. 

Our investigation yields one set of archival data obtained from the Panoramic Survey Telescope and Rapid Response System (Pan-STARRS). Around T$_0$, Pan-STARRS covered the host galaxy of the magnetar for a total period of about one year, from {November 04}, 2012 to November 18, 2013, with a total observational time of {912} seconds\cite{Panstarr_survey2016}. Those observations are listed in Extended Data Table \ref{tab:pan_obs}, and are available as single-epoch ``warp" images in Pan-STARRS1 data archive, which are astrometrically and photometrically calibrated\footnote{\url{https://outerspace.stsci.edu/display/PANSTARRS/PS1+Warp+images}}. No variable point source was found within a circular region of {3$^\prime$} radius from the galaxy center in those observations. For each warp image, we calculate the 3$\sigma$ upper limit of a merger-type point source detection using the following procedure:

\begin{enumerate}
\item For each pixel in the warp image, we calculate its A-B magnitude following the guide as listed in Pan-STARRS document\footnote{\url{https://outerspace.stsci.edu/display/PANSTARRS/PS1+Stack+images}} and Ref\cite{2020ApJS..251....4W, 2020ApJS..251....5M} as 
\begin{equation}
m_{AB}=A-2.5 \times \mathrm{log} C + 2.5\times \mathrm{log} t_{exp},
\end{equation}
where {A is the magnitude zero point,} C is the photon counts in each pixel, $t_{exp}$ is the exposure time in units of second. 

\item {The K half-light radius is $R_{50}$ = 4.94$^{\prime \prime}$\cite{2006AJ....131.1163S}.} Since the magnetar is likely a merger event, we assume it occurred at the outskirt of the galaxy {with the normalized offset $r_{\rm off}$ with a similar value as in the case of GRB 050724\cite{2010ApJ...708....9F}, namely $r_{\rm off}$ = $\frac{R_{\rm off}}{R_{\rm gal}}$ = 0.68. This allows us to place the magnetar at a distance of $R_{\rm off}$ = 3.36$^{\prime \prime}$ from the galaxy center.}

\item With an same inclination angle of the host galaxy, we selecte a region of an {ellipse annulus with the outer semi-major axis of 4.30$^{\prime \prime}$ and the inner semi-major axis of 3.19$^{\prime \prime}$}, as shown in Extended Data Figure \ref{fig:mag}. We then calculate the magnitude of each pixel within the annulus and plot their probability density distribution, as shown in Extended Data Figure \ref{fig:mag}. The 3$\sigma$ value of such distribution is used to determine the 3$\sigma$ upper limit of a point source at radius $R_{\rm off}$.

\end{enumerate}

Through the above steps, we obtained the 3$\sigma$ upper limit of a point source at $r$ {for a total of 15 warp images}, as listed in Extended Data Table 3 and plotted in Figure 4. We note that 8 warp images in Extended Data Table 3 are considered ``bad" (e.g., the magnitude value of each pixel is marked as ``nan" due to gaps between detectors, bad pixel regions, etc\footnote{https://outerspace.stsci.edu/display/PANSTARRS/PS1+Image+Cutout+Service}.) so their upper limits are unavailable.

\section*{Possible optical light curves of the progenitor}

The progenitor of the magnetar can be either a collapsar-type GRB/SN, a superluminous supernova (SLSN), or a neutron star merger event. All those cases are associated with significant optical emission. Moreover, if the progenitor is an off-axis GRB, the off-axis afterglow could last for weeks. Below we test if any of the expected optical emission is consistent with the upper limits obtained from the Pan-STARR data.

For the case of SLSN, we choose SN2015bn as a representative and assume that it could serve as the progenitor of the magnetar in this study. SN 2015bn was located in a faint host galaxy at D$_L\simeq$ 544.8 Mpc\cite{Modjaz2016A_2015bn}. The optical data are available at the Open Supernova Catalog\footnote{https://sne.space}. We first fit the multi-wavelength optical light curves using the magnetar-powered model described in Ref.\cite{Nicholl2015b} and obtain the physical parameters to describe the observational data of SN 2015bn. As listed in Extended Data Table 4, after being updated on the luminosity distance and the magnetic field using those values of the hyper flare GRB 130310A, the parameter set is utilized to calculate the $r$-band magnitude using the {same} model. The resulted light curve is plotted as a solid green line in Figure 4. One can see the SLSN magnitude is way above the PAN-STARR upper limit. The weeks-scale coverage of PAN-STARR observations around $T_0$ can rule out the existence of an SLSN, 

For the case of a collapsar-type GRB/SN, We choose SN 1998bw as an representative. SN 1998bw was located in a barred spiral galaxy (ESO 184-G82) at D$_L\simeq$ 37 Mpc\cite{Guillochon2017_SNCatlog}. The data are also available in the Open Supernova Catalog. After fitting the multi-wavelength optical light curves using the $^{56}$Ni-decay model\cite{Nadyozhin1994_Ic}, the best-fit physical parameters expect for the luminosity distance (as listed in Extended Data Table 4) are used to calculate the $r$-band Type Ic SN light curve at $70$ Mpc. The resulting light curve, as shown in Figure 4, is well above the PAN-STARR upper limit. We can thus also rule out the classical collapsar-type supernova origin of the magentar.

For the case of a neutron star merger, one should consider the existence of a kilonova (KN). A comparable case is the observed AT2017gfo with GRB 170817A. AT2017gfo was located in an elliptical galaxy (NGC 4993) at $D_L\simeq$ 40 Mpc\cite{Andreoni2017}. The optical data are available at the Open Kilonova Catalog\footnote{https://kilonova.space}. We apply the model employed in Ref\cite{Metzger2019_kilonova} and fit the multi-wavelength optical light curves to obtain the best-fit physical parameters as listed in Extended Data Table 4. After updating the luminosity distance, we plot the $r$-band kilonova light curve at 70 Mpc in Figure \ref{fig:optical}. Our results suggest that the peak magnitude of the kilonova could be above the upper limit of point-source detection. On the other hand, the theoretical KN light curve fell in the observational gap of PAN-STARR. So the neutron star merger origin of the magnetar is allowed.

Giving the fact that no GRB was found between $\sim$ $T_0-\tau_{c,max}$ and $T_0$ and that a kilonova could not be ruled out, the putative short GRB associated with the merger event was likely off-axis with respect to the observer, unless it was missed during the 35.4\% off-FOV-time of Fermi/GBM. We compared the off-axis afterglow of GRB 170817A with the upper limit constrained from PAN-STARR observations. To do so, we first fit the $r$-band afterglow light curve of GRB 170817A using the structured jet model in Ref.\cite{Troja2020_off_axis_afterglow}, which gave us a set of best-fit physical parameters, as listed in Extended Data Table 4. After replacing the luminosity distance with $70$ Mpc, we calculated the predicted off-axis afterglow in our case. Such a light curve is plotted in Figure 4. One can see that the off-axis afterglow of the short GRB is undetectable by PAN-STARR. We also calculated the on-axis short GRB afterglow for the same set of parameters at $70$ Mpc and found that such an afterglow can also survive the observational constraints if it occurred in the observational gaps. Considering the 35.4\% off time for GRB, this suggests that an on-axis short GRB and afterglow as the magnetat progenitor is not ruled out. However, in view that the event rate density of off-axis neutron star merger events is much higher than that of the on-axis events, we regard an off-beam neutron star merger as a more likely progenitor of the magnetar.

\section*{Late-time afterglow and multiwavelength counterparts}

The late-time afterglow of the event should undergo the slow-cooling phase. The flux density $F_\nu$ at frequency $\nu$ follows $F_{\nu}\propto \nu^{-p/2}t^{(2-3p)/4}$. Using the following typical parameters in a standard forward shock model:
\begin{enumerate}
 \item electron distribution power law index, $p$ = 2.2
 \item energy fraction in electrons, $\epsilon_e$ = 0.1
 \item energy fraction in magnetic fields, $\epsilon_B$ =0.01
 \item accelerated electron fraction, $\xi_N$ =1.0
 \item ISM density $n_0$ = 1.0 $\rm cm^{-3}$
 \end{enumerate}
we can calculate the flux densities in X-ray at $10^{18}\rm~Hz$ and optical band at $10^{14}\rm~Hz$ at $t\sim $ 8.5 years as \cite{2002ApJ...568..820G}: 
\begin{align}
f_{\nu}(X) \simeq 1.21\times 10^{-10}{\rm mJy}~\epsilon_{\rm e, 0.1}^{p-1} \epsilon_{\rm B, 0.01}^{(p-2)/4} E_{\rm iso, 49}^{(2+p)/4} t_{8.5 \rm yrs}^{(2-3p)/4} d_{27}^{-2} \nu_{18}^{-p/2},
\end{align}
and 
\begin{align}
f_{\nu}(\rm O) \simeq 3.04\times 10^{-6}{\rm mJy}~\epsilon_{\rm e, 0.1}^{p-1} \epsilon_{\rm B, 0.01}^{(p-2)/4} E_{\rm iso, 49}^{(2+p)/4} t_{8.5 \rm yrs}^{(2-3p)/4} d_{27}^{-2} \nu_{14}^{-p/2}.
\end{align}

Magentasrs are also believed to be the energy source of Fast Radio Bursts (FRBs). For a typical FRB with luminosity of {$10^{42}$} erg/s, its observed flux density at 70 Mpc is about $50$ Jy in typical raido bands, which is detected by the ground-based radio telescopes, such as CHIME, MeeKat, FaST and VLA.

The luminosity of the X-ray bursts of the Galactic SGRs are typically {$\sim 10^{35}$} erg/s. Putting it at 70 Mpc, we found that it too faint for current X-ray missions including Chandra and Swift.

\end{methods}


\section*{Data Availability}
 Processed data are presented in the tables and figures in the paper. Source and optical observational data are available upon reasonable requests to the corresponding authors. The {\it Fermi}/GBM data are publicly available at \url{https://heasarc.gsfc.nasa.gov/FTP/fermi/data}.
 
\section*{Code Availability}
Upon reasonable request, the code (mostly in {\it Python}) used to produce the results and figures will be provided.

%


\end{bibunit}

\begin{addendum}
\item 
B.B.Z acknowledges support by the National Key Research and Development Programs of China (2018YFA0404204), the National Natural Science Foundation of China (Grant Nos. 11833003, U2038105, 12121003), the science research grants from the China Manned Space Project with NO.CMS-CSST-2021-B11, and the Program for Innovative Talents, Entrepreneur in Jiangsu. We also acknowledge the use of public data from the Fermi Science Support Center (FSSC). BBZ thank Rongfeng Shen for helpful discussions. ZJZ acknowledges the helpful discussions and constructive comments from Ken Chen, Jianbin Weng, Ke Xu, Shengyu Yan and Prof. Lang Shao.

 \item[Author Contributions] 
BBZ and YSY initiated the study. BBZ, ZJZ, JSW and BZ coordinated the scientific investigations of the event. BBZ, ZKL, YHY, JHZ, XIW, and JY processed and analyzed the data. YCM helped on independently verifying the periodic signal identification. BBZ wrote the paper with the contributions from all coauthors.

 \item[Competing Interests] The authors declare that they have no competing financial interests.
 
 \item[Additional information]
 \ \\
 \textbf{Correspondence and requests for materials} should be addressed to B.-B. Z.

\end{addendum}

\begin{extendeddata}

\label{fig:table1}
\begin{table*}
\begin{center}
\caption{\textbf{Observational properties of the hyper flare event GRB 130310A}. The total fluence and peak flux are calculated in 10--10,000 keV energy band. All errors correspond to the 1$\sigma$ credible intervals.}
\label{tab:para}
\begin{tabular}{ll}
\hline
\hline

Observed Properties & GRB 130310A\\
\hline
Abrupt rise time & $\sim 20$ ms \\
Steep decay time & $\sim 900$ ms \\
$T_{\rm 90}$ (burst) & $2.90_{-0.22}^{+0.31}$ s\\
$T_{\rm 90}$ (precursor) &$1.30_{-0.44}^{+0.57}$ s\\
Duration of the ``sharp peak" & {$\sim$ 27 ms}\\
Waiting time between precursor and main burst &$\sim 3.79$s \\
$\alpha$ at peak &$-1.15_{-0.02}^{+0.26}$ \\
$E_{\rm p}$ at peak & $10.75_{-1.52}^{+3.25}$ MeV\\
Time-integrated $\alpha$ (main burst )& $-1.10_{-0.01}^{+0.01}$ \\
Time-integrated $E_{\rm p}$(main burst) & $2.73_{-0.20}^{+0.23}$ MeV\\
{Time-integrated $\alpha$ } (sharp peak )& $-1.19\pm 0.02$ \\
{Time-integrated $E_{\rm p}$(sharp peak) }& $9.80_{-1.22}^{+2.23}$ MeV \\
Total fluence & $4.12_{-0.32}^{+0.32}\times10^{-5}$ $\rm erg\ cm^{-2}$\\
Peak flux& $4.06_{-0.44}^{+0.44}\times 10^{-4}\ \rm {erg\ cm^{-2}\ s^{-1}}$ \\
Possible host galaxy & g0927191-170053 \\
Luminosity distance & 69.41 Mpc (z=0.0155) \\
Isotropic energy (precursor) $E_{\rm\gamma,iso}$ & $6.30_{-3.75}^{+7.49}\times 10^{46}\rm\ erg$\\
Isotropic energy (sharp peak) $E_{\rm\gamma,iso}$ & $5.61_{-0.46}^{+0.51}\times 10^{48}\rm\ erg$\\
Isotropic energy (main burst) $E_{\rm\gamma,iso}$ & $2.37_{-0.18}^{+0.18}\times 10^{49}\rm\ erg$\\
Peak luminosity $L_{\rm\gamma,p,iso}$ & $2.30_{-0.25}^{+0.25}\times 10^{50}\rm\ erg\ s^{-1}$\\
Average tail luminosity $L_{\rm tail}$ & $5.63\times 10^{48}\rm\ erg\ s^{-1}$\\
Spectral lag & $\sim$ 0 s\\
Minimal variability time scale $\delta t$ & 0.0075 s \\
{f parameter} & $11.22\pm 1.32$ \\
\hline
\hline
\end{tabular}
\end{center}
\end{table*}

\begin{table*}
\begin{center}
\caption{Spectral properties of GRB 130310A}
\label{exttab2}
\begin{tabular}{cccccccc}
\hline
\hline
\multirow{2}{*}{Component} &\multirow{2}{*}{$t_1$} & \multirow{2}{*}{$t_2$} & \multirow{2}{*}{Model} & \multicolumn{2}{c}{Parameter(s)}& \multirow{2}{*}{PGSTAT/dof} \\
\cline{5-6}
 & & & &$\alpha$ & $E_{\rm p}$ or KT (keV) &&\\
\hline
Precursor& -0.57 & 0.35 & BB & ... & $45.06_{-5.69}^{+13.57}$ & 307.2/349 \\
\hline
Sharp Peak & 4.108 & 4.135 &CPL& $-1.19_{-0.02}^{+0.02}$ & $9802.23_{-1220.55}^{+2234.88}$ & 354.3/349 \\
\hline
\hline

\multirow{9}{*}{Main Burst} & 3.85& 4.05&\multirow{9}{*}{CPL}&$-2.09_{-0.40}^{+2.31}$ & $858.50_{-438.85}^{+1668.71}$& 163.3/349\\
&4.05& 4.118&&$-1.29_{-0.06}^{+0.04}$ & $3943.29_{-801.26}^{+2681.62}$ & 270.2/349 \\
&4.118& 4.135&&$-1.15_{-0.03}^{+0.02}$ & $10749.48_{-1522.56}^{+3247.12}$& 356.3/349\\
&4.135& 4.155&&$-1.04_{-0.04}^{+0.03}$ & $4964.09_{-807.51}^{+1233.92}$ & 306.5/349 \\
&4.155& 4.185&&$-0.73_{-0.08}^{+0.06}$ & $1153.87_{-155.79}^{+262.58}$ & 275.7/349 \\
&4.185& 4.215&&$-0.79_{-0.09}^{+0.06}$ & $940.61_{-111.29}^{+279.29}$ & 251.1/349 \\
&4.215& 4.265&&$-0.49_{-0.10}^{+0.10}$ & $593.03_{-58.23}^{+106.63}$ & 279.2/349 \\
&4.265& 4.485&&$-0.69_{-0.06}^{+0.05}$ & $706.63_{-63.63}^{+104.73}$ & 355.6/349 \\
&4.485& 5.25&&$-1.07_{-0.05}^{+0.04}$ & $885.18_{-127.37}^{+247.26}$ & 316.8/349 \\
&4.05& 5.25&& $-1.10_{-0.01}^{+0.01}$ & $2732.60_{-197.80}^{+229.75}$ & 421.2/349 \\
\hline
\hline
\end{tabular}
\end{center}
\end{table*}

\begin{table*}[]
 \centering
 \begin{tabular}{cccccc}
 \hline
 \hline
 Date&Number of observations & Observation time &Exposure time (s) & Filters& Upper limit for a point source detection (m$_{AB}$) \\
 \hline
 2012-11-04&1&14:46:23.563&30 & y& 18.72\\ 
 \hline
 2012-11-25&2&15:28:02.527&30&z& Unconstrained\\
 & &15:43:20.493&30&z& Unconstrained\\
 \hline
 2012-12-28&2&13:36:32.675&45&i& 18.98\\
 & &13:54:38.657&45&i& 18.94\\
 \hline
 2012-12-29&2&14:09:08.155&40&r& 19.49\\
 & &14:26:01.300&40&r& 19.48\\ 
 \hline
 2013-01-25&4&10:54:12.288&40&r&Unconstrained\\
 & &10:55:56.784& 40&r&19.38\\
 & &11:11:00.673& 40&r&Unconstrained\\
 & &11:12:44.948& 40&r&19.38\\
 \hline
 2013-01-26&2&11:30:33.887&45&i& 19.06\\
 & &11:49:06.794&45&i& 19.11\\ 
 \hline
 2013-04-01&4&06:44:26.925&43&g&Unconstrained\\
 & &06:51:45.444& 43&g&20.87\\
 & &07:02:04.029& 43&g&Unconstrained\\
 & &07:09:21.358& 43&g&20.75\\
 \hline
 2013-04-24&5&05:32:38.923&30&y&Unconstrained\\
 & &05:47:32.901& 30&y&Unconstrained\\
 & &05:54:35.440& 30&y&18.91\\
 & &06:11:14.971& 30&z&18.97\\
 & &06:26:05.774& 30&z&18.91\\
 \hline
 2013-11-18&1&15:52:03.043& 80& y&18.45\\
 \hline
 \hline
 \end{tabular}
 \caption{\textbf{List of PAN-STARR observations that covered the host galaxy around $T_0$. } The up-limit calculation is detailed in Methods.}
 \label{tab:pan_obs}
\end{table*}

\begin{table*}
\begin{tiny}

\begin{center}
\label{tab:nova_par}
\begin{tabular}{ll|ll|ll|ll}
\hline
\hline

SLSN & Value & Ic-SN & Value & Kilonova & Value & Off-axis afterglow & Value \\
\hline
Surface field& $B_{s}$ $\sim 10^{15.5}$ G&Nickel fraction & $f_{\rm Ni}$ = 0.023 &Blue ejecta mass &$M_{\rm ej}^{\rm blue}$ = 0.023 $M_\odot$ & fraction of magnetic energy & $\varepsilon_B$ = $10^{-4}$\\
NS mass &$M_{\rm NS} $ = 1.79 $M_\odot$&Opacity to $\gamma$-rays &$\kappa_{\gamma} $ = 0.11 ${\rm cm}^{2} {\rm g}^{-1}$ &Blue ejecta velocity &$v_{\rm ej}^{\rm blue} \sim$ 78000 $\rm km\ s^{-1}$ &fraction of electron population energy& $\varepsilon_e$ = 0.04\\
Spin period &$P_{\rm spin} $ = 1.60 $\rm ms$&Ejecta mass & $M_{\rm ej} $ = 1.78 $M_\odot$ &Blue opacity &$\kappa_{\rm ej}^{\rm blue}$ = 0.5 ${\rm cm}^{2} {\rm g}^{-1}$&power-law slope of the electron population & $p$ = 2.17\\
Optical opacity& $\kappa $ = 0.20 ${\rm cm}^{2} {\rm g}^{-1}$&Host H number density &$n_{\rm H, host}$ $\sim$ $10^{17.8} {\rm cm}^{-3}$&Blue temperature&$T^{\rm blue}$ = 3983 K&ISM density&$n_0$ = 0.002 $\rm cm^{-3}$\\
Opacity to $\gamma$-rays &$\kappa_{\gamma} $ = 0.01 ${\rm cm}^{2} {\rm g}^{-1}$&Photosphere temperature &$T_{\rm min}$ = 6607 K&Red ejecta mass &$M_{\rm ej}^{\rm red}$ = 0.050 $M_\odot$&jet orientation&$\theta_{V}$ = 0.40 rad \\
Ejecta mass &$M_{\rm ej} $ = 0.1 $M_{\odot}$& Explosion time& $t_{\rm exp}$ = -54.52 days&Red ejecta velocity& $v_{\rm ej}^{\rm red} \sim 44700\ \rm km\ s^{-1}$&jet core width&$\theta_c$ = 0.07 rad \\
Ejecta velocity &$v_{\rm ej} $ $\sim 5460\ \rm km\ s^{-1}$&Variance &$\sigma$ = 0.38 mag&Red temperature &$T^{\rm red}$ = 3745 K&jet total width&$\theta_w$ = 0.47 rad \\
Host galaxy Extinction &$A_V = 0.08$ mag&Ejecta velocity & $v_{\rm ej}$ $\sim$ 77625 $\rm km\ s^{-1}$ &Red opacity &$\kappa_{\rm ej}^{\rm red}$ = 10 ${\rm cm}^{2} {\rm g}^{-1}$&\\
Temperature &T = $ 8.32\times10^{13}$ K & Luminosity distance & $\simeq$ 70 Mpc & Variance &$\sigma$ = 0.24 mag \\
Variance &$\sigma$ = 0.18 mag& & &Luminosity distance & $\simeq$ 70 Mpc\\
Luminosity distance & $\simeq$ 70 Mpc & & & & \\
 & & & & & \\
 & & & & & \\
\hline
\hline
\end{tabular}
\caption{\textbf{Model parameters of the SLSN, Ic-SN, kilonova and off-axis afterglow}.}
\end{center}
\end{tiny}

\end{table*}

\begin{figure*}

 \centering
 \includegraphics[width=0.7\textwidth]{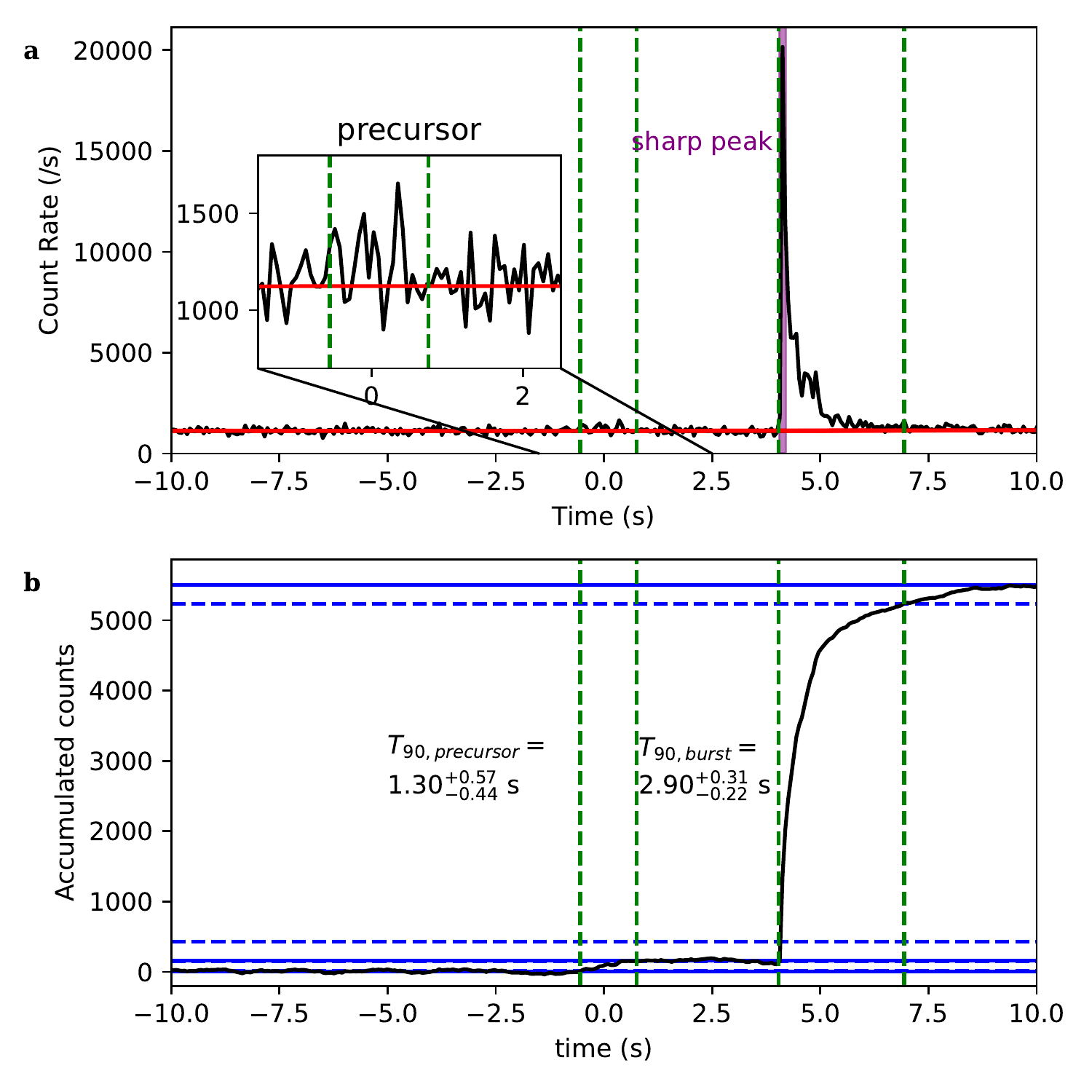}
 \caption{
 \noindent {Light curve and duration of GRB 130310A.} \textbf{a}, the black solid lines show the light curve obtained by the GBM-{na} data in energy range of 8-1000 KeV. The red solid line represents the level of background. The sharp spike (SP) is at $\sim 4s$. and marked with purple shaded area. The inset brackets the precursor region. \textbf{b,} black line represents the accumulated count light curve. The blue horizontal dashed (solid) lines are plotted at 5$\%$ (0$\%$) and 95$\%$ (100$\%$) of the total accumulated counts, respectively. The two regions marked by the green vertical dashed lines are the $T_{90}$ intervals of the precursor and the main burst.}
 \label{fig:t90lc}
\end{figure*}

\begin{figure*}[!htb]
\centering
\includegraphics[width=0.85\textwidth]{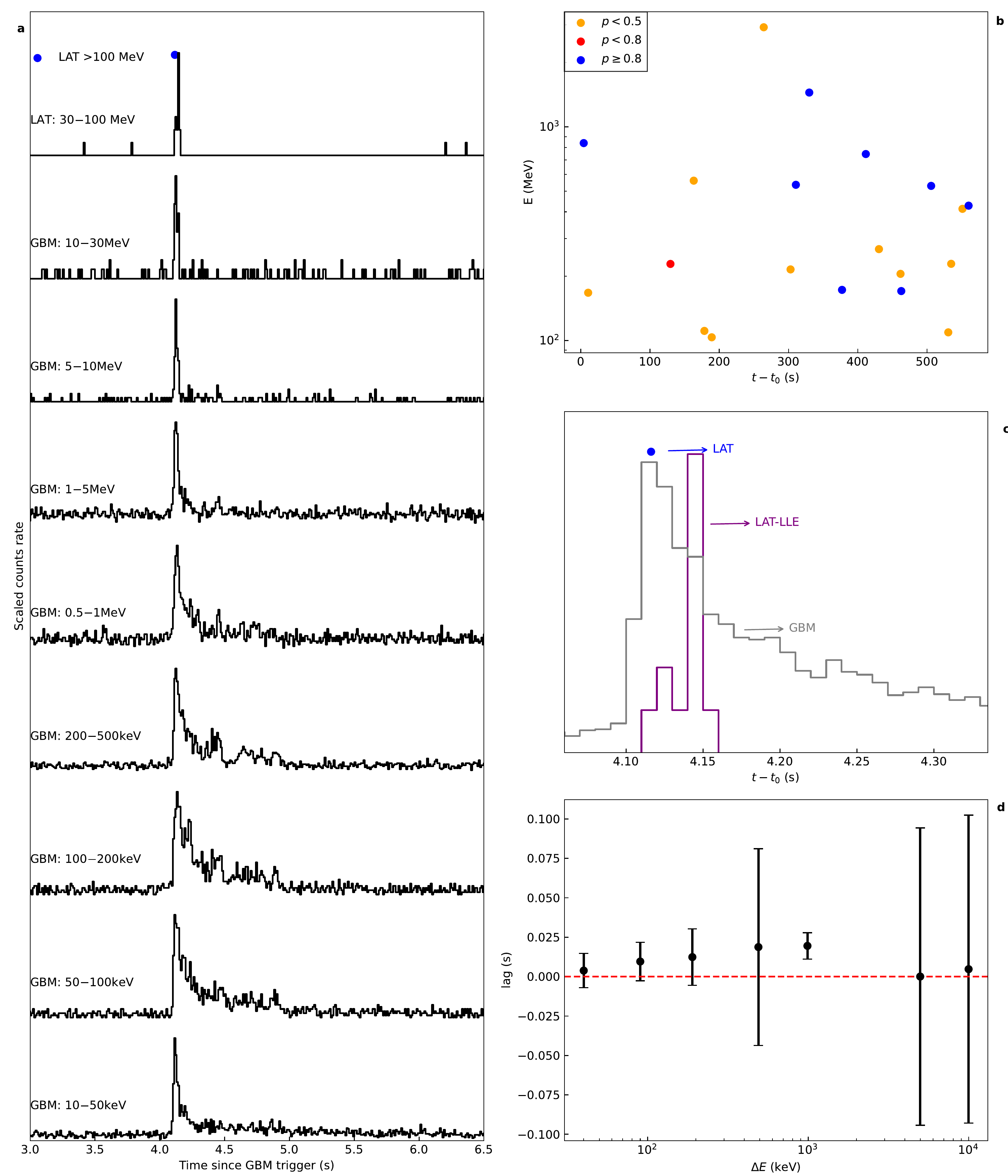}
\caption{\noindent \textbf{Multi-wavelength light curves, high-energy photon detection and time .} \textbf{a,} Multi-wavelength light curves detected with GBM and LAT. The filled blue circle is an event detected by LAT around the peak time of the GBM light curve. \textbf{b,} LAT events with energy above 100 MeV in $0-800$ s. Different colors represent different ranges of probabilities that the photon count may originate from this event. \textbf{c,} Comparison between LAT-LLE and GBM light curves. \textbf{d,} Energy dependent lags between the lowest energy(10-20 keV) band and any higher energy bands. All error bars represent 1$\sigma$ uncertainties.}
\label{edf2}
\end{figure*}

\begin{figure*}[!htb]
\centering 
\includegraphics[width=1.0\textwidth]{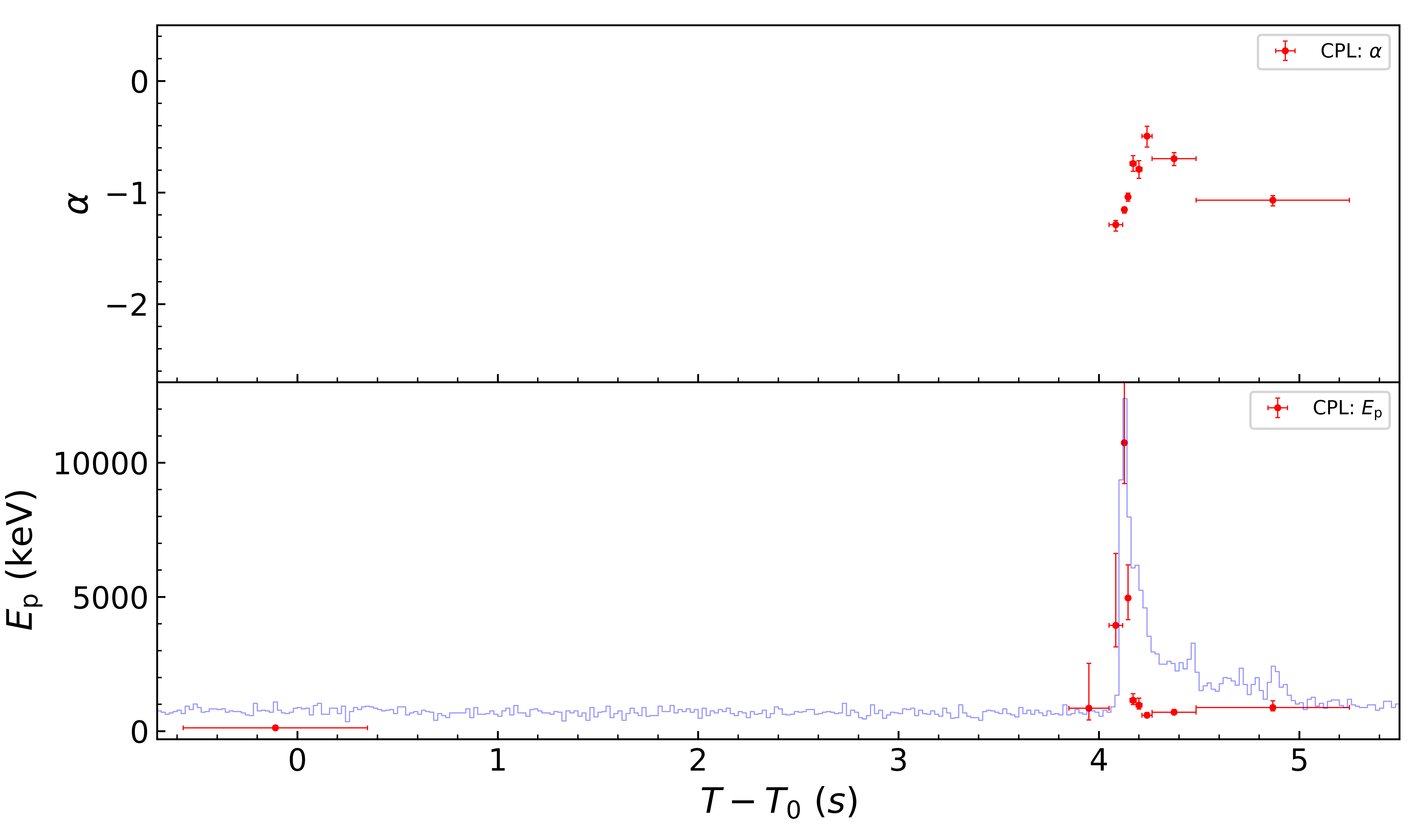}
\caption{\noindent \textbf{Spectral evolution of the GRB 130310A. } The time-resolved spectra are fitted with a CPL model for the main burst and a black body model for the precursor. The top panel shows the evolution of the low-energy photon index, $\alpha$ for the main burst. The bottom panel shows the evolution spectral peak, $E_{\rm p}$, for both the precursor and the main burst, where an equivalent \textcolor{black}{$E_{\rm p} \sim 2.82$ kT} (Ref. \cite{Lu2012ApJ...758L..34Z}) is adapted for the precursor. All error bars represent 1-$\sigma$ uncertainties}
\label{edf3}
\end{figure*}

\begin{figure*}

\centering
 \includegraphics[width=0.47\textwidth]{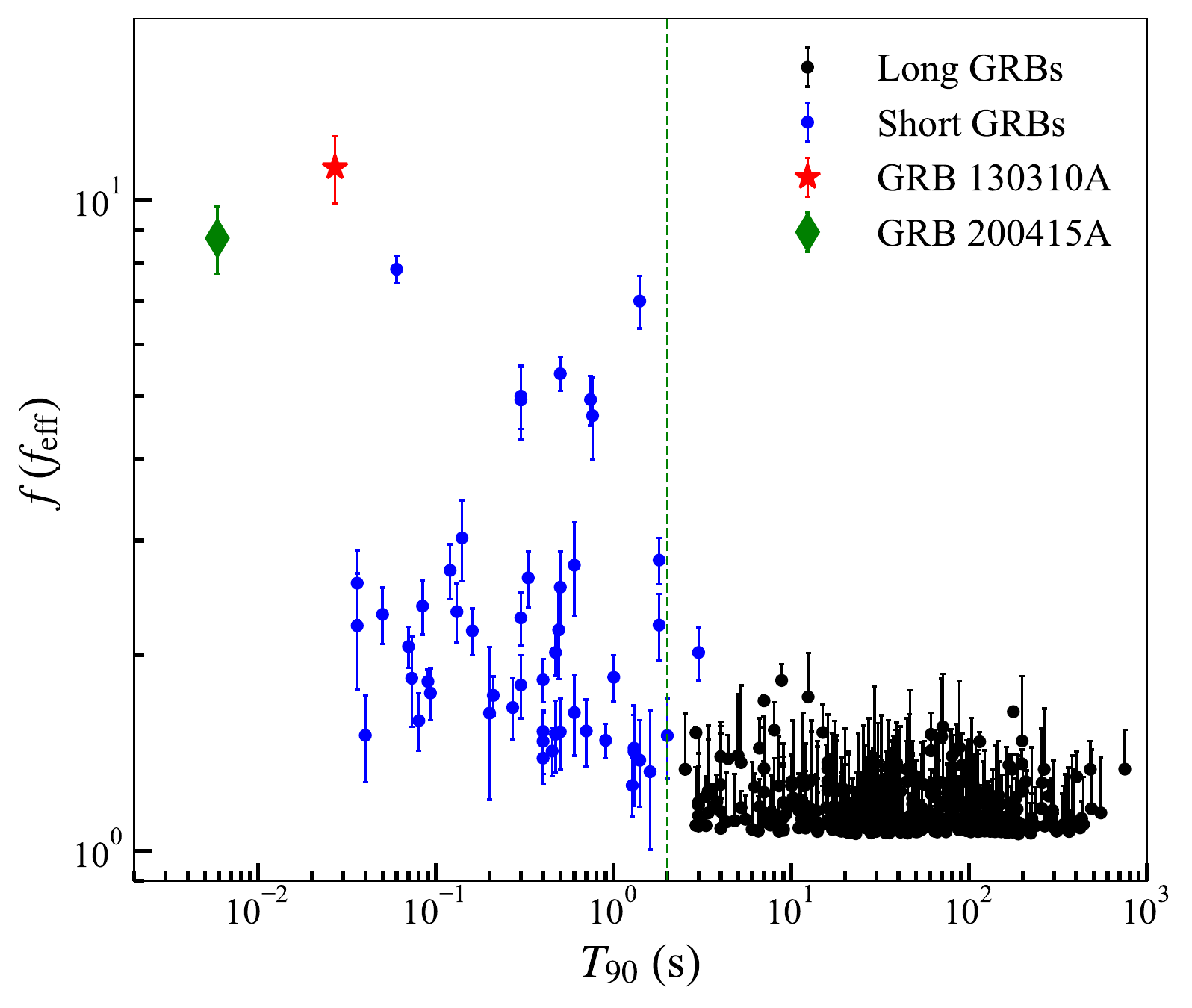}
 \caption{$T_{90}-f (f_{\rm eff})$ plot. The blue circles represent the $f$ value of short GRBs, and black circles represent the $f_{\rm eff}$ of long GRBs. The effective amplitude parameter $f_{\rm eff}=\frac{F_{\rm p'}}{F_{\rm b}}$ for a long GRB is measured by scaling down the burst until ``pseudo GRB'' is shorter than 2 s (Ref.\cite{Lv14}) .The green vertical dashed line is the boundary of 2 s. The green diamond represents the MGF GRB 200415A. The red-star-marker highlights the sharp peak of GRB 130310A .}
 \label{fig:f_para}
\end{figure*}

\begin{figure*}
\centering
 \includegraphics[width=0.8\textwidth]{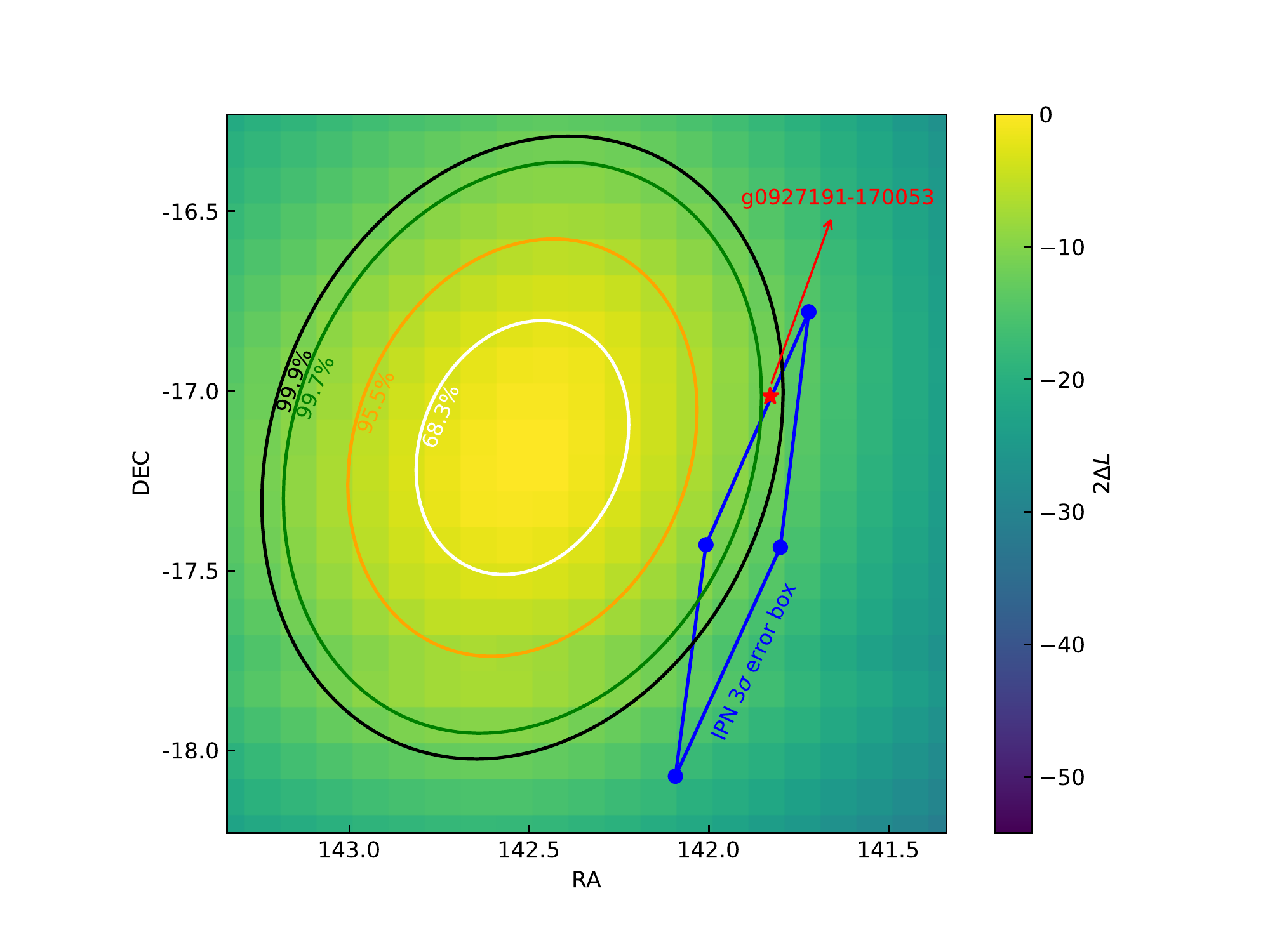}
 \caption{\textbf{Localization Test Statistics (LTS) map of the hyper flare GRB 130310A obtained with LAT data in the time range of 0$-$800 s.} The white, orange, green and black counters are the 68.3\%, 99.5\%, 99.7\% and 99.9\% confidence regions, respectively. $3\sigma$ IPN error box\cite{2013GCN.14284....1G} (blue parallelogram) and the potential host galaxy g0927191-170053 (red star) are over-plotted on top of the LAT error circles.}
 \label{fig:latloc}
\end{figure*}

\begin{figure*}
\vspace{3.0cm}
 \hspace{0cm}
\subfigure{
\begin{minipage}[t]{0.3\linewidth}
\includegraphics[width=1.5\linewidth]{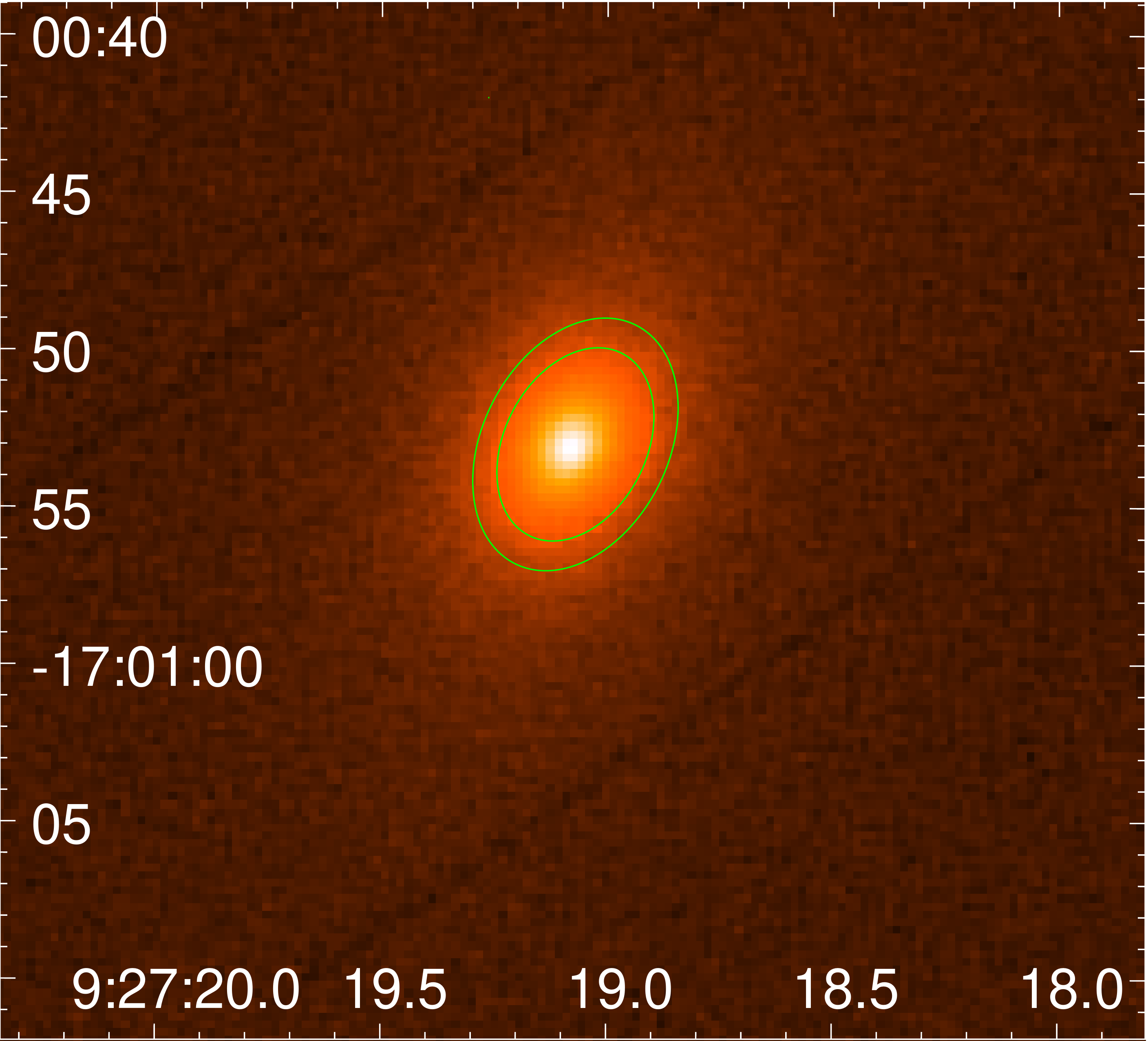}
\put(-240,200){\bf a}
\end{minipage}%
}%
\hspace{3.0cm}
\subfigure{
\begin{minipage}[t]{0.36\linewidth}
\includegraphics[width=1.5\linewidth]{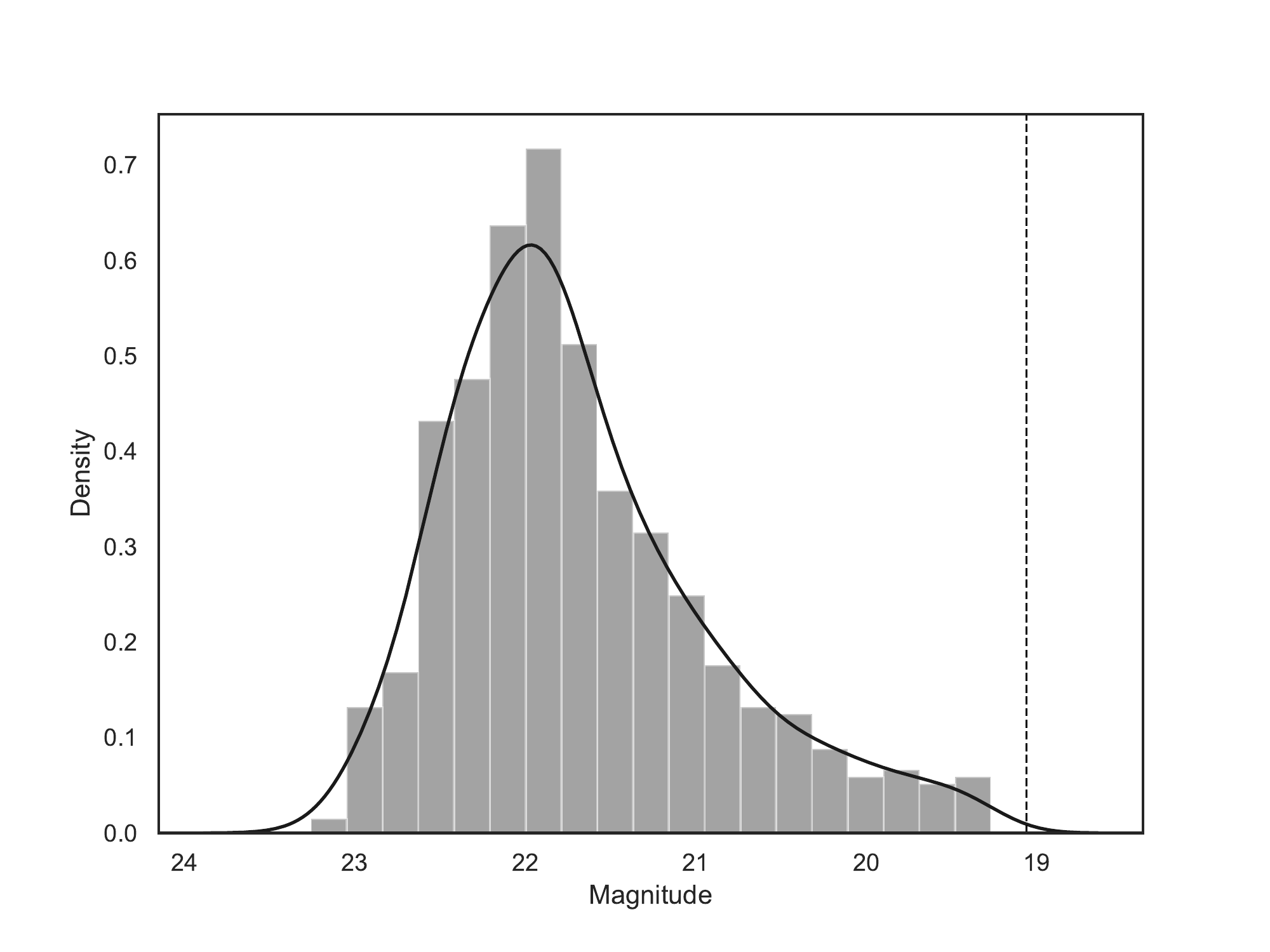}
\put(-265,174){\bf b}
\end{minipage}%
}%
\caption{\textbf{An example illustrating the method of the up-limit calculation. }{\bf a,} A warp image of the host galaxy, g0927191-170053, obtained from the PAN-STARR archive for the observation performed on 2012-12-28T13:36:32.675. The green ellipse annuli is the region to measure the up-limit of the point-source transient. {\bf b,} The magnitude distribution of each pixel within the annuli. The black solid line is the probability density function, and the black dashed line marks the 3$\sigma$ level.}
\vspace{1cm}
\label{fig:mag}
\end{figure*}

\begin{figure*}

\centering
 \includegraphics[width=0.8\textwidth]{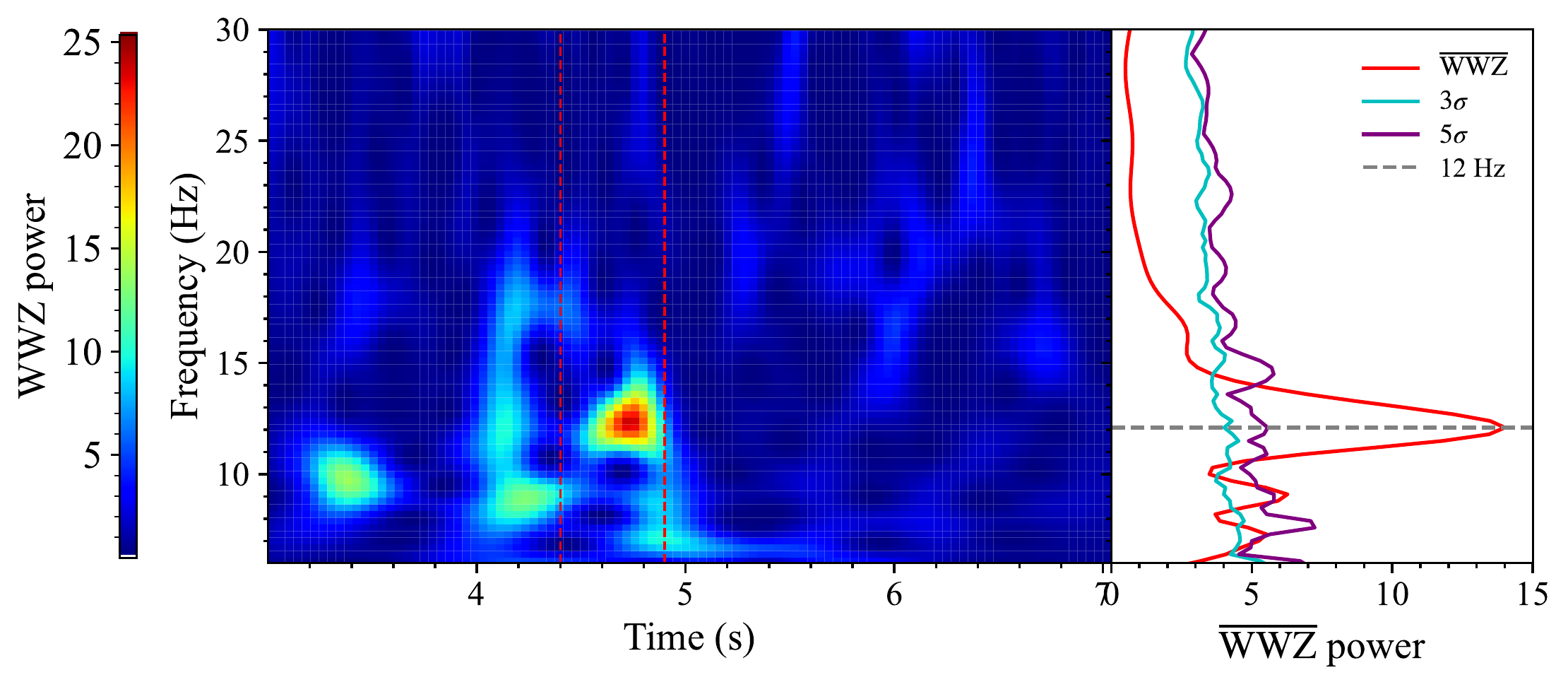}
 \caption{\textbf{The wavelet plot of the detrended data. A significant signal is shown at 12 HZ between 4.4 and 5.0 s.} The left plot is derived from WWZ (Weighted Wavelet Z-Transform) transform of the detrended data. The right panel shows the mean wwz power for different frequencies.}
 \label{fig:specgram}
\end{figure*}

\begin{figure*}
\centering
\begin{tabular}{ll}
\textbf{a} & \textbf{b} \\
\includegraphics[width=0.43\textwidth]{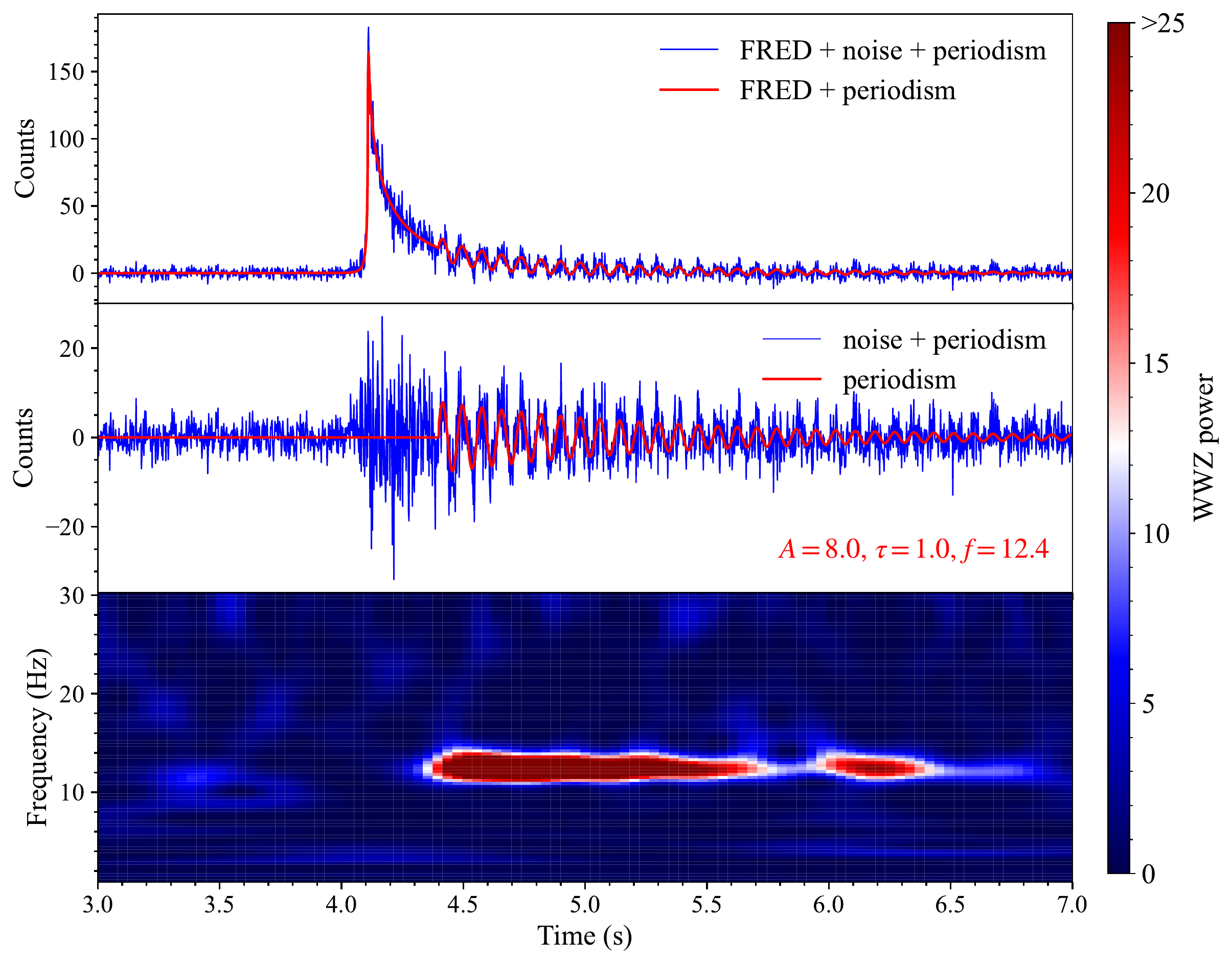} &
\includegraphics[width=0.43\textwidth]{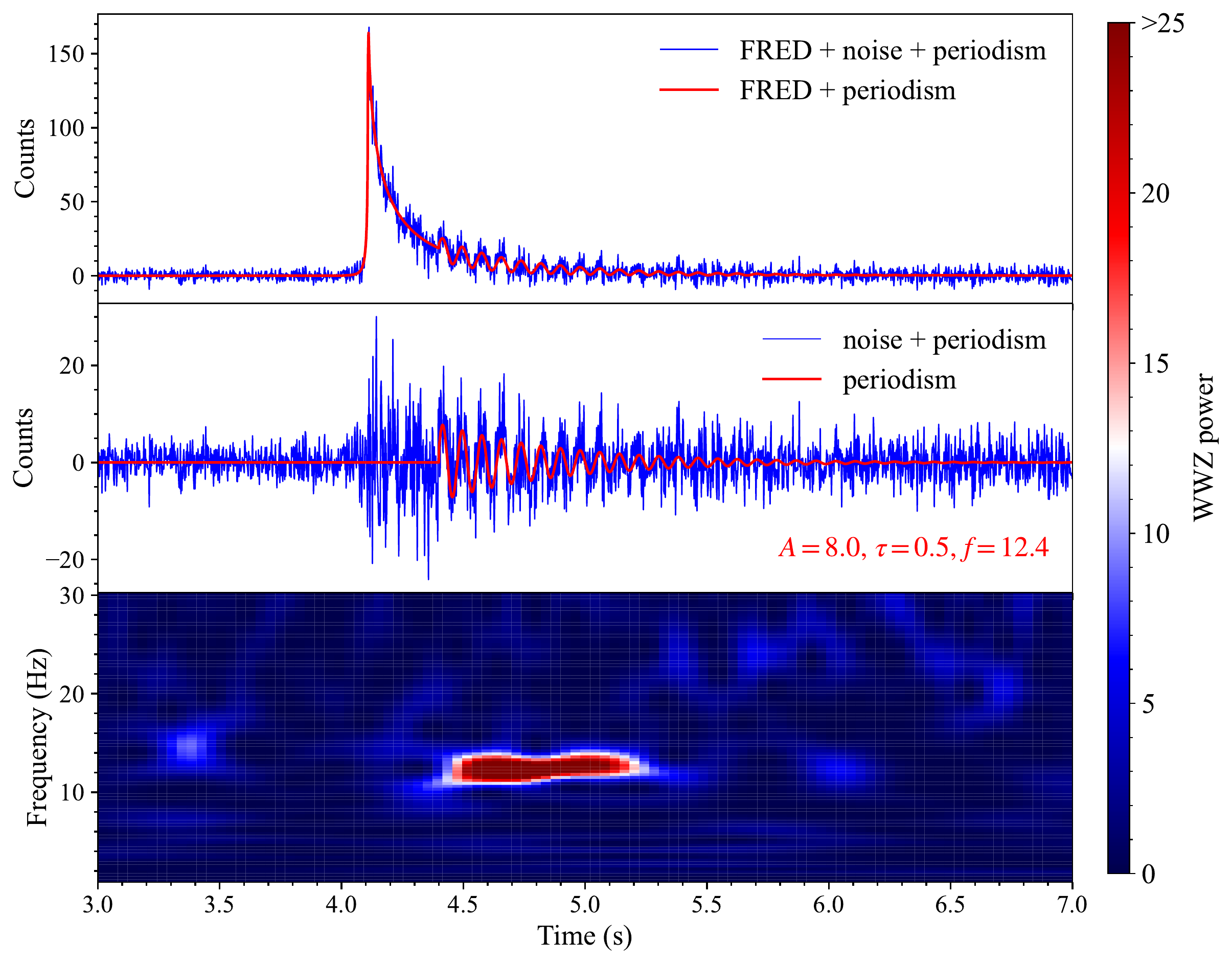} \\ \textbf{c} & \textbf{d} \\
\includegraphics[width=0.43\textwidth]{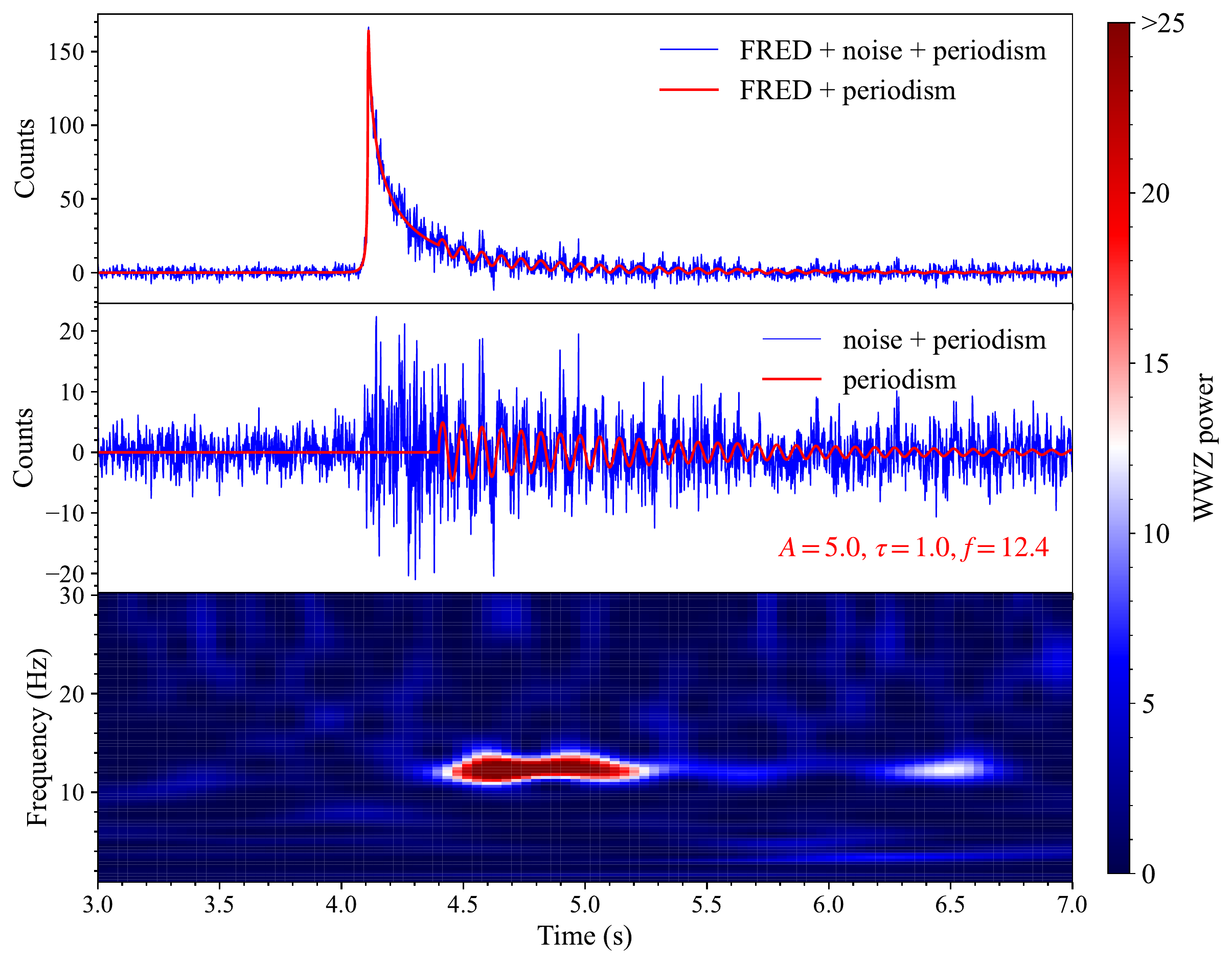} & \includegraphics[width=0.43\textwidth]{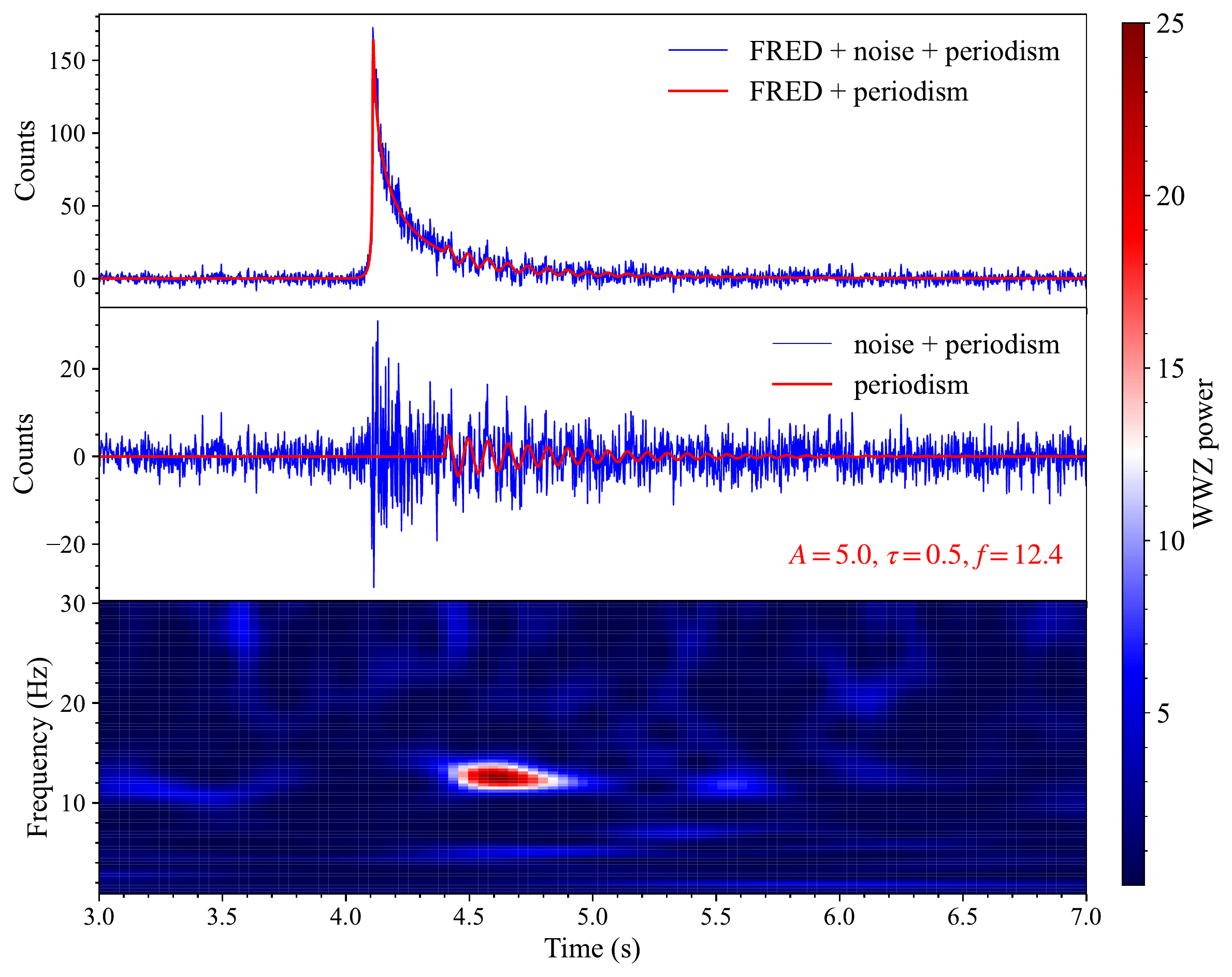} \\
\textbf{e} & \textbf{f} \\ 
\includegraphics[width=0.43\textwidth]{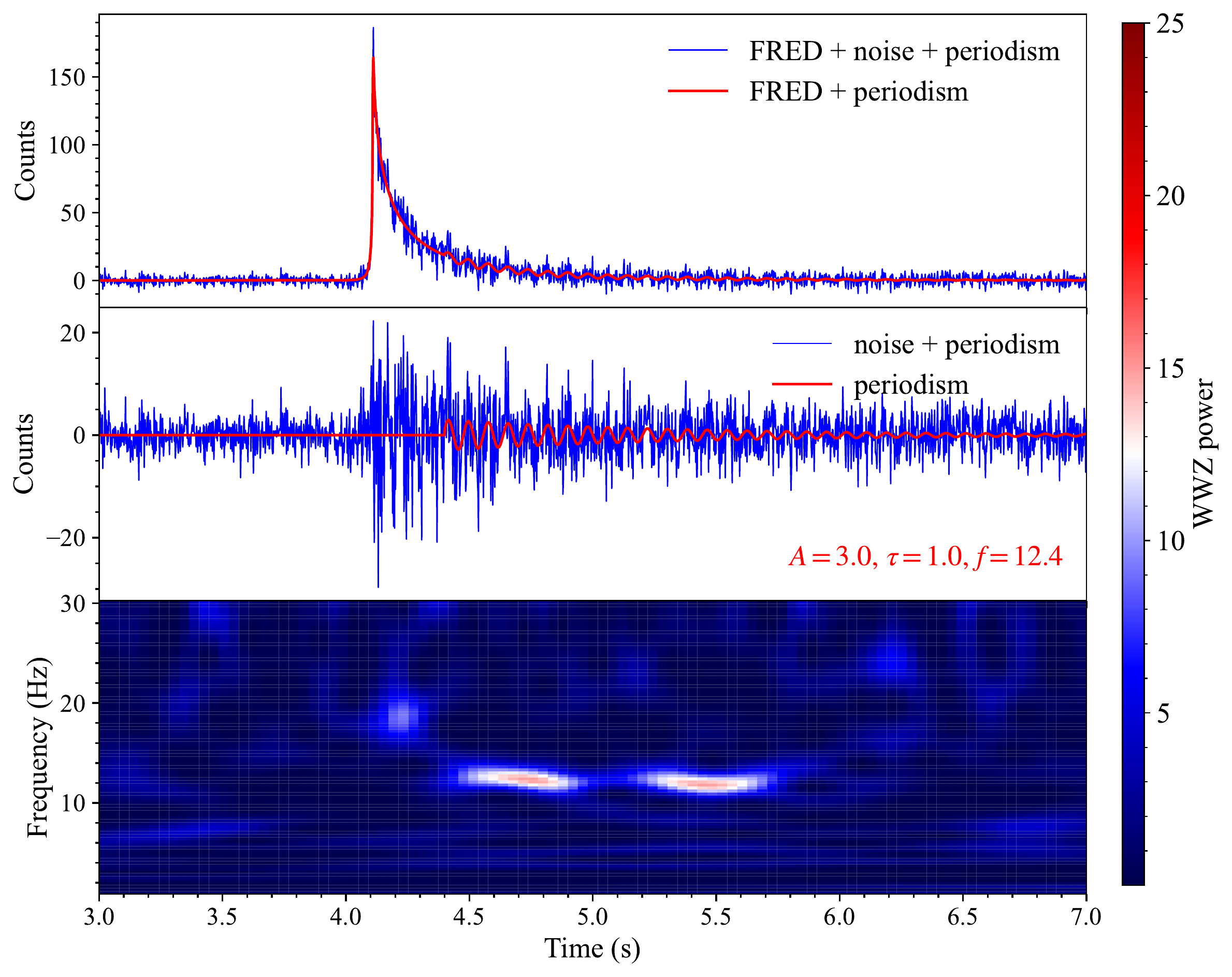} & \includegraphics[width=0.43\textwidth]{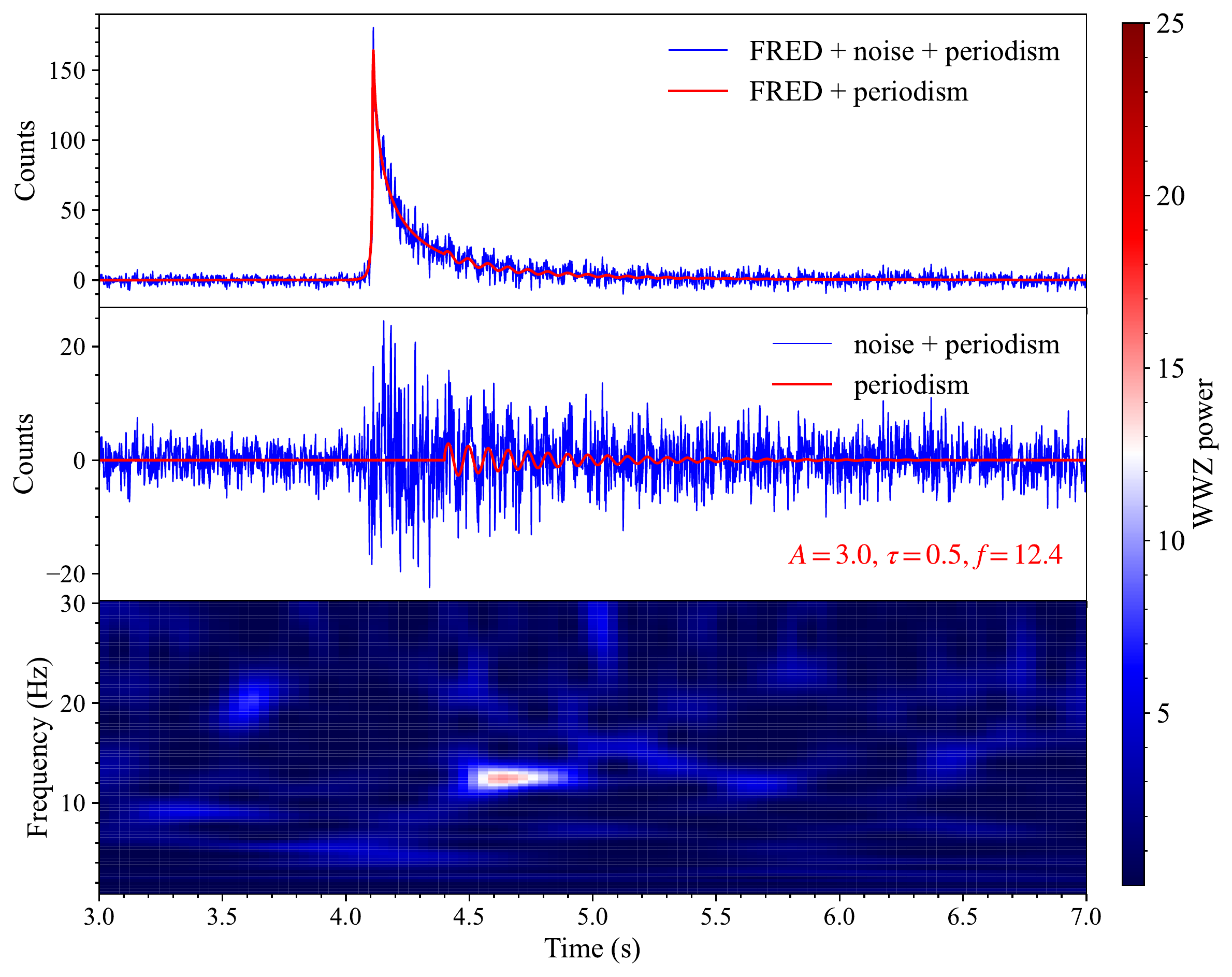} \\
\end{tabular}
\caption{\textbf{An illustration shows that the periodic signal can become undetectable depending on its modulation strength to the observed flux.} In each plot, the simulated light curve is composed of a FRED-shaped pulse (solid red line on the top panel), a random noise signal (blue in the middle panel), and a 12.4-Hz periodic signal applied between 4.4 s and 7.0 s (solid red line in middle panel). With the strength of the periodic signal decreasing from \textbf{a} to \textbf{f} (see Methods), one can see the detectable range of the 12-Hz period becomes more concentrated to its maximum-strength region. The detecting range of the periodic signal in \textbf{d} is similar to that of GRB 130310A. }
\end{figure*}

\end{extendeddata}

\end{document}